\documentclass[acmsmall,screen]{acmart}

\usepackage{algorithmic}
\usepackage{graphicx}
\usepackage{textcomp}
\usepackage{xcolor}

\usepackage{hyperref}
\usepackage{enumitem}
\usepackage{tabularx}
\usepackage{textcomp}
\usepackage{tcolorbox}
\usepackage{amsmath}
\usepackage{amsfonts}
\usepackage{booktabs}
\usepackage{makecell}
\usepackage{tabularx}
\usepackage{booktabs}
\usepackage{pgf-pie}
\usepackage{multirow}
\usepackage{xspace}
\usepackage{balance}
\usepackage{siunitx}
\usepackage{array}
\usepackage{ragged2e}

\newcommand{\ppathf}{{\sc PPatHF}\xspace}
\newcommand{\fixmorph}{{\sc FixMorph}\xspace}
\newcommand{\skyport}{{\sc SkyPort}\xspace}
\newcommand{\mystique}{{\sc Mystique}\xspace}

\newcommand{\tabmargin}{\vspace{-2mm}} 

\newcommand{\boxmargin}{1mm}
\newtcolorbox{myboxc}{
    colback=gray!15!white,
    arc = 0pt, outer arc = 0pt,
    boxsep=0pt, left = 3pt, right = 0pt, top = 0pt, bottom = 0pt, 
    leftrule=3pt, bottomrule=0pt,toprule=0pt, rightrule=0pt,
    left = \boxmargin, right = \boxmargin, top = \boxmargin, bottom = \boxmargin
}

\AtBeginDocument{%
  }

\setcopyright{acmlicensed}
\copyrightyear{2018}
\acmYear{2018}
\acmDOI{XXXXXXX.XXXXXXX}
\acmConference[Conference acronym 'XX]{Make sure to enter the correct
  conference title from your rights confirmation email}{June 03--05,
  2018}{Woodstock, NY}
\acmISBN{978-1-4503-XXXX-X/2018/06}

\begin{document}

\title{BackportBench: A Multilingual Benchmark for Automated Backporting of Patches}

\author{Zhiqing Zhong}
\authornote{Both authors contributed equally to this research.}
\email{zhiqingzhong@link.cuhk.edu.cn}
\orcid{0000-0003-2304-9619}
\affiliation{%
  \institution{The Chinese University of Hong Kong, Shenzhen (CUHK-Shenzhen)}
  \country{China}
}

\author{Jiaming Huang}
\authornotemark[1]
\affiliation{%
  \institution{The Chinese University of Hong Kong, Shenzhen (CUHK-Shenzhen)}
  \country{China}}
\email{jiaminghuang@link.cuhk.edu.cn}
\orcid{0000-0003-0798-1633}

\author{Pinjia He}
\authornote{Corresponding author.}
\affiliation{%
  \institution{The Chinese University of Hong Kong, Shenzhen (CUHK-Shenzhen)}
  \country{China}}
\email{hepinjia@cuhk.edu.cn}
\orcid{0000-0003-3377-8129}






\renewcommand{\shortauthors}{Trovato et al.}

\begin{abstract}
Many modern software projects evolve rapidly to incorporate new features and security patches.
It is important for users to update their dependencies to safer versions, but many still use older, vulnerable package versions because upgrading can be difficult and may break their existing codebase.
Software developers can mitigate this problem by backporting security patches to older releases. However, manually backporting is time-consuming and error-prone. The effectiveness of existing automated backporting techniques on general software remains unclear since they typically target only code-hunk or function-level patch porting scenarios and are evaluated with imperfect metrics.

To facilitate the development and evaluation of automated backporting techniques, we introduce BackportBench, the first comprehensive benchmark suite for patch backporting problem.
BackportBench is a multilingual benchmark that contains 202 patch backporting problems from PyPI, Maven, and npm, each with executable Docker environments and relevant test cases.
We evaluated existing patch porting methods and LLM-based techniques that have the potential to adapt to this task using BackportBench. The results show that the agentic method has outperformed traditional patch porting methods, especially on cases that require logical and structural changes. However, the performance varies across different programming languages.
Based on the findings, we draw several implications for researchers and software practitioners in future work on automated backporting.
\end{abstract}


\keywords{Do, Not, Use, This, Code, Put, the, Correct, Terms, for,
  Your, Paper}

\received{20 February 2007}
\received[revised]{12 March 2009}
\received[accepted]{5 June 2009}

\maketitle

\section{Introduction}

Popular software projects often maintain multiple branches to balance introducing new features with preserving compatibility for existing users~\cite{wingerd1998high,shihab2012effect,tan2022understanding}. As software evolves, vulnerabilities may emerge, affecting multiple branches and putting their users at risk. 
Although developers eventually fix these vulnerabilities, patches are often released only for the latest branch, leaving older but still widely used branches unpatched~\cite{decan2021back}. To remove such vulnerabilities, users must upgrade to a patched version in the latest branch, which can require significant effort to address breaking changes~\cite{pashchenko2020qualitative, jayasuriya2023understanding, venturini2023depended}. 
To alleviate this challenge, patch backporting is a helpful practice that adapts patches for other versions and makes it easier to deliver patches to users on older branches. 
However, manual backporting is error-prone and time-consuming, often leading to delays or leaving some branches unpatched ~\cite{tan2022understanding,rodriguez2015increasing,chakroborti2022backports,wu2025mystique}. Therefore, automated patch backporting is recommended in this situation.

Researchers have proposed several patch porting techniques in recent years~\cite{shariffdeen2020automated,shariffdeen2021automated,shi2022backporting,yang2023enhancing,pan2024automating,wu2025mystique}. 
Early work relied mainly on predefined patch patterns to backport patches~\cite{shariffdeen2021automated,yang2023enhancing}.
With the emergence of large language models (LLMs), recent researchers have explored the effectiveness of LLMs on porting patches. For instance,
Pan et al.~\cite{pan2024automating} proposed \ppathf, which employs a fine-tuned LLM to adapt patches from functions that have been stripped of irrelevant context to hard fork projects.
Wu et al.~\cite{wu2025mystique} introduced \mystique, an approach that extracts semantic and syntactic signatures to guide a fine-tuned LLM to port patches.
Despite these advances, the effectiveness of current patch porting methods on general software packages from open-source software (OSS) ecosystems (e.g., PyPI, Maven, and npm) remains unclear for the following limitations:
\begin{itemize}[leftmargin=*]
\item \textbf{Limitation 1: Existing methods focus on porting patches within local code hunks or functions.} Recent research shows that a large portion of backporting involves handling incompatibilities across different files~\cite{chakroborti2024study}. This suggests that focusing on local code hunks or functions during patch porting is unrealistic, since it curtails the relevant information across files and simplifies the problem.
\item \textbf{Limitation 2: Ad hoc designs reduce generalizability across programming languages and vulnerability types.} Some approaches rely on ad hoc designs, such as leveraging unique features provided by Clang/LLVM to transform the source code~\cite{shariffdeen2021automated}, or curating sink functions to represent vulnerable logic in PHP injection vulnerabilities~\cite{shi2022backporting}. While these methods work in certain cases, they are tied to specific languages or vulnerability categories, limiting their broader applicability.
\item \textbf{Limitation 3: Equivalence-based and similarity-based metrics are unstable and not comprehensive for evaluating backporting effectiveness.} Most prior work evaluates backported patches by checking whether they are syntactically and semantically equivalent to the human-written ground truth. This process typically requires manual inspection~\cite{wu2025mystique, shariffdeen2021automated}, making it labor-intensive, unstable across independent evaluations, and not comprehensive for nonequivalent patches that are actually effective. When such equivalence cannot be achieved, some prior work turns to adopt similarity-based metrics (e.g., edit distance, and CodeBLEU~\cite{ren2020codebleu}) to measure how similar the backported patches are to the human-written ground truth, which is less effective~\cite{ray2012case}.
We argue that whether backported patches achieve equivalence is not the only way to prove the patch is effective. In many cases, requiring equivalence is too strict, especially when patches require significant adaptation. This perspective is also common in other coding-related tasks~\cite{dong2025survey,he2025llm,li2025cocoevo}.
\end{itemize}

To alleviate these limitations and facilitate the development and evaluation of automated backporting techniques for a more practical scenario, we introduce BackportBench, the first comprehensive benchmark suite for the patch backporting problem.
BackportBench contains 202 real-world vulnerability patch backporting task instances curated from three popular OSS ecosystems, PyPI , Maven, and npm, covering Python, Java, and JavaScript, respectively. 
Consider a situation where a vulnerability affects two versions of a software package, but only one version is patched. BackportBench defines the backporting problem as generating a patch for the unpatched software version based on the existing patch and the code differences between the two versions.
This formulation shifts the patch porting problem from code-hunk level or function level to repository level, offering a more realistic representation of software maintenance challenges.
To enable stable and comprehensive evaluation, each task instance provides an executable Docker environment with scripts for running relevant test cases, thereby aligns with the criteria for a successful benchmark~\cite{bhuiyan2023secbench}.

Furthermore, we design three research questions to understand how are high-quality backports distributed in the real-world software development, and the potential of existing automated patch porting techniques~\cite{wu2025mystique,pan2024automating} and GitHub issue resolution techniques~\cite{xia2024agentless,yang2024swe} on BackportBench.
We found that agentic method has outperformed traditional patch porting methods. However, the performance varies across different programming languages.

In summary, we make the following contributions:

\begin{itemize}[topsep=5pt]
\item We introduce BackportBench, the first multilingual patch backporting benchmark that includes executable environments and test cases for validation. The BackportBench is available on \url{https://github.com/BackportBench/BackportBench}.
\item We analyze the distribution of high-quality patch backports in real-world software development.
\item We conduct evaluations for not only current patch porting methods, but also LLM-based techniques that have potential to adapt to this task.
\item We provide an in-depth discussion of future directions for patch backporting techniques and more applications of BackportBench.
\end{itemize}

\section{Related Work}
\subsection{Patch Backporting and Transplantation}
\label{sec:tra_baselines}
Previous studies have unveiled that patch backporting and transplantation are helpful for package users to eliminate security risks in dependencies without adopting their software to a newer major release, but the delay of ported patches and the infrequency of backporting practice significantly weaken this benefit~\cite{decan2021back, pan2024automating, shariffdeen2021automated}.
Therefore, several automated patch backporting techniques were proposed.
\fixmorph~\cite{shariffdeen2021automated} backports patches in Linux kernel to its lower version by synthesizing syntactical transformation rules.
TSBPORT~\cite{yang2023enhancing} backports patches in Linux Kernel by first locating the code hunks to be backported with the program dependency graph, then applying patch migration based on the fine-grained patch type.
\texttt{PatchWeave}~\cite{shariffdeen2020automated} uses failing test cases and concolic execution to guide patch transplantation between C/C++ libraries.
\skyport~\cite{shi2022backporting} focuses on backporting injection vulnerability patches in PHP web frameworks by defining and leveraging a list of sink functions for this type of vulnerability.
\ppathf~\cite{pan2024automating} ports patches from source project to its hard fork by removing less relevant content in input function and uses a fine-tuned LLM to adapt the patch to the hard fork.
\mystique~\cite{wu2025mystique} extracts semantic and syntactic function signatures to guide a fine-tuned LLM to port patches in C and Java projects.

However, existing methods face several challenges. First, the design of many of the methods relies on language-specific features, which restricts their applicability to particular languages or even to certain types of vulnerability patches.
Second, these methods focus on porting patches within local code hunks or functions. Nevertheless, recent research shows that a large portion of backports involves handling incompatibilities across different files~\cite{chakroborti2024study}.
Third and most importantly, most of these methods evaluate results based on syntactical and semantical equivalence. This typically requires manual inspection, making it difficult to be consistent across independent evaluation in different papers, and labor-instensive. When such equivalence cannot be achieved for difficult cases, some works turn to adopt similarity-based metrics (e.g., edit distance, and CodeBLEU~\cite{ren2020codebleu}) to measure how similar the backported patch is to human-written ground truth, which is less effective~\cite{ray2012case}.

\subsection{Repository-level Benchmark for GitHub Issue Resolution}
While automated code generation~\cite{galenson2014codehint, chen2021evaluating, zhuo2024bigcodebench} and automated program repair~\cite{monperrus2018automatic, jiang2018shaping, xu2024aligning} have been long-standing research problems in software engineering, previous benchmarks usually curtail the code context of these problems for simplicity~\cite{jimenez2023swe}.
Recently, several benchmarks have been proposed to evaluate performance on resolving repository-level coding problems.
Jimenez et al.~\cite{jimenez2023swe} proposed SWE-bench, the first GitHub issue resolution benchmark designed to provide a broader set of real-world challenges beyond closed-form code completion. Each challenge in SWE-bench comprises a codebase along with an issue description. Resolving such issues requires editing the codebase so that the related test suite passes. 
OpenAI released SWE-bench Verified~\cite{openai2024swe}, which is a human-validated subset of SWE-bench that enables more reliable evaluation.
Subsequently, to extend such evaluation beyond Python, SWE-bench-java~\cite{zan2024swe} and Multi-SWE-bench~\cite{zan2025multi} were introduced, covering 7 additional popular programming languages.
Guo et al.~\cite{guo2025omnigirl} further introduced OmniGIRL to challenge LLMs with issues whose descriptions include images, collected from repositories across different domains.
Nevertheless, existing repository-level benchmarks do not cover patch backporting issues, which typically require understanding two versions of a codebase and the original patch.

\subsection{LLM-based Methods for GitHub Issue Resolution}
Jimenez et al.~\cite{jimenez2023swe} initially investigated how effective LLMs are at solving GitHub issues by integrating retrieval methods to select files to include in the prompt context. However, even when combined with BM25 retrieval~\cite{robertson2009probabilistic} or Oracle retrieval (i.e., directly providing the actual files edited in the ground truth), none of the best-performance models at that time solved more than 5\% of the instances in SWE-bench.
With the rise of LLM agent applications for other software engineering problems~\cite{feldt2023towards, liu2024large, wang2025agents, zhang2024large}, researchers have begun exploring various management strategies for using LLMs to resolve GitHub issues in SWE-bench~\cite{yang2024swe, zhang2024autocoderover,antoniades2024swe,xia2024agentless,yang2025enhancing,tao2024magis,su2025learn,wang2024openhands,gao2025trae}.
Following recent surveys~\cite{yang2025survey,dong2025survey}, these methods can be grouped into two categories: agentic methods and procedural methods.
Agentic methods usually assign decision-making to LLM agents, which devise plans and execute actions to resolve issues.
Typical examples include SWE-agent~\cite{yang2024swe}, which integrates tools that allow the LLM to interact with the coding environment;
AutoCodeRover~\cite{zhang2024autocoderover}, which improves localization of edited code with a set of AST-based search tools;
and SWE-Search~\cite{antoniades2024swe}, a multi-agent framework that incorporates Monte Carlo Tree Search into the agent control loop to improve decisions.
Procedural methods require the LLM to follow a fixed pipeline.
For example, Agentless~\cite{xia2024agentless} performs hierarchical localization for edited code, then generates multiple patches, and selects one using test cases.
KGCompass~\cite{yang2025enhancing} builds a knowledge graph from the codebase, repository issues, and pull requests to facilitate localization, and then proceeds to patch generation and selection.
Overall, agentic methods emphasize autonomous decision-making, whereas procedural methods emphasize expert-designed workflows.

\section{BackportBench Construction}
\label{sec:Methods}
\begin{figure}[t]
    \centering
    \includegraphics[width=1\linewidth]{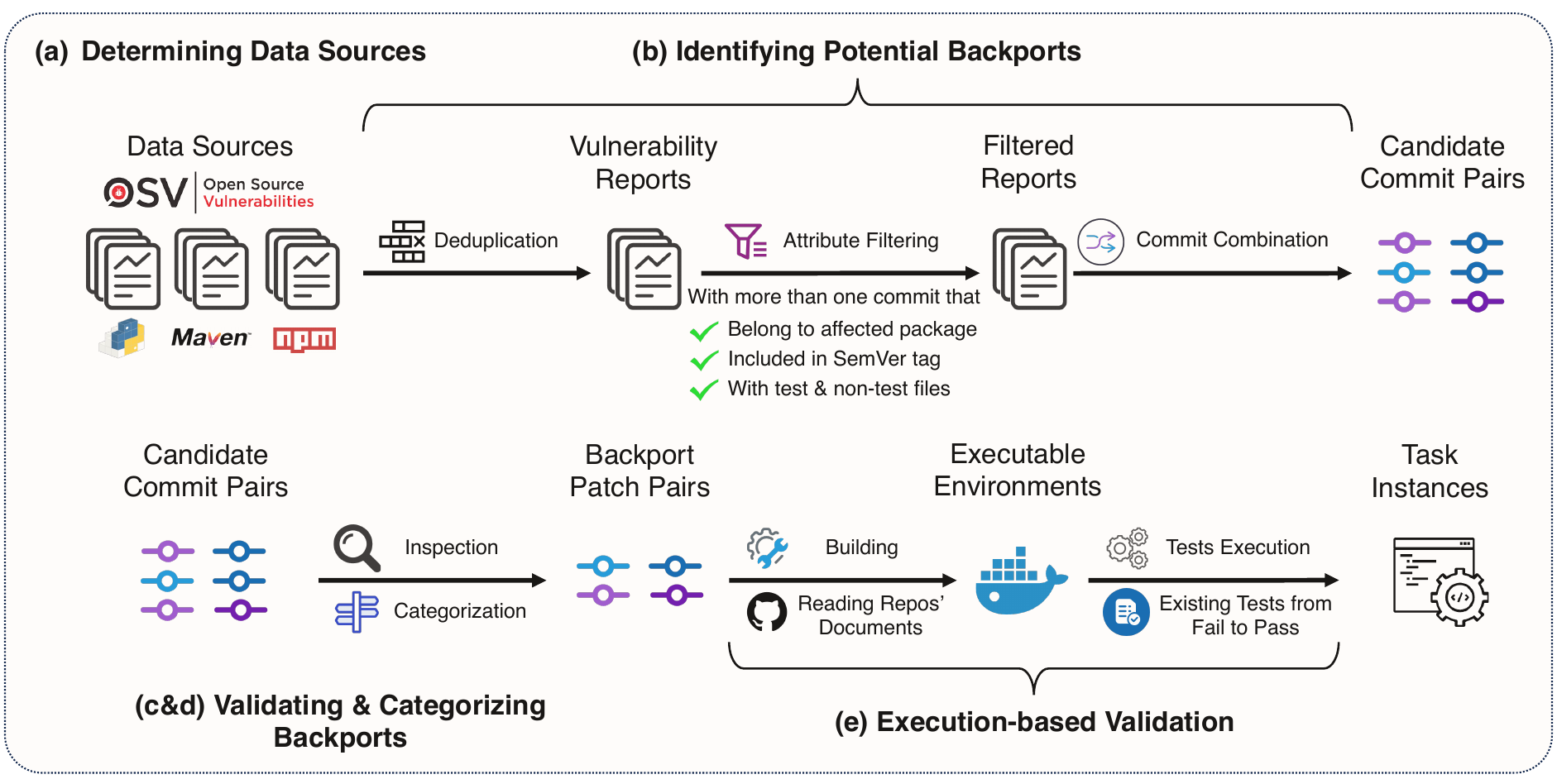}
    \caption{Overview of BackportBench construction.}
    \label{fig:workflow}
\end{figure}

Figure~\ref{fig:workflow} illustrates the overview of our methodology for constructing BackportBench.
First, we chose the OSV database~\cite{OSV} and its referenced GitHub commits as data sources, since they allow us to mine backporting instances at scale (Section~\ref{sec:data_sources}). 
Next, we identified potential backporting instances from the above data sources. Specifically, a backporting instance is defined as a pair of commits that patch the same vulnerability in different versions of codebase (Section~\ref{sec:identifying_data}).
Subsequently, we further confirmed whether the commit pairs truly represent a backporting relationship (Section~\ref{sec:validating}) and categorized them based on the level of adaptation required for the backported patches (Section~\ref{sec:categorize}).
Finally, to ensure the confirmed backporting instances are usable for evaluation, we built Docker environments for the backported codebases and validated the quality of their test suites through execution (Section~\ref{sec:building_env}).

\subsection{Determining Data Sources}
\label{sec:data_sources}
The first challenge to develop BackportBench is finding backporting task instances at scale. 
Since backporting is an expensive maintenance practice, even many popular repositories do not backport patches to other widely used major versions~\cite{decan2021back, zhang2023mitigating}.
Fortunately, we also observe that repositories are more likely to backport patches that address severe issues (e.g., vulnerabilities)~\cite{chakroborti2022backports}.
Based on this observation, we focus on vulnerability patches and use OSV database~\cite{OSV} to identify them. OSV is an open source, multi-ecosystem database that is widely adopted by previous research~\cite{pan2024towards,zahan2023software,yang2025tracing}, aggregating vulnerabilities from National Vulnerability Database (NVD)~\cite{NVD}, GitHub Advisory Database~\cite{GAD}, and other sources.
Specifically, we focus on vulnerability records from the PyPI, Maven, and npm ecosystems in the 2024.8.29 OSV data dump, where each vulnerability record may include GitHub commit links in its references. These commits may indicate vulnerability patches or their backported patches.
We further describe our procedure for identifying backporting data in the next section.

\subsection{Identifying Potential Backporting Data}
\label{sec:identifying_data}
The OSV vulnerability records and their linked GitHub commits are noisy, as the records may be duplicated, and the commits may be unrelated to patch backporting.
This duplication occurs when OSV aggregates records for the vulnerability from multiple sources. Although each vulnerability record is assigned a unique ID, OSV records include an \texttt{aliases} field that links duplicate records~\cite{Chang2025aliases}.
To deduplicate, we group all records linked via \texttt{aliases} as representing the same vulnerability. For each group, we take the union of all GitHub commit links in their references to identify potential backporting data.

From the union of GitHub commits for each vulnerability, we create high-quality commit pairs for further validations.
In BackportBench, each backporting task instance consists of two patches: the original patch and its backported patch. Therefore, we first exclude vulnerabilities that have fewer than two GitHub commits in the union.
However, some commits may be irrelevant or of low quality. We filter these out and only keep deduplicated vulnerability records with more than one valid commit that meet the following criteria.
First, \textbf{the commit should belong to the affected package in the vulnerability record}. Some commits for the dependencies of the affected packages may be included in the references for more detailed information about the vulnerability. Therefore, we filter out GitHub commits that do not belong to the affected packages based on the repository name in the commit and the affected package name in the vulnerability record. 
Second, \textbf{the commit should be included in a semantic versioning commit tag}. Unlike related benchmarks~\cite{zan2025multi, hu2024real, mundler2024swt,nashid2025issue2test} that only consider code merged into the main branch as accepted by the repository's maintainers, backported patches are often distributed on different branches, which may even be deleted after the release. In this scenario, if a commit is either tagged with a semantic version on GitHub, or is an ancestor of such a tagged commit, this can also indicate that it has been accepted and released by the maintainers.
Last but not least, \textbf{the commit should include changes to both test files and non-test files}. This criterion allows us to perform execution-based evaluations for the backporting task. Following previous repository-level benchmarks~\cite{jimenez2023swe, guo2025omnigirl}, we consider files that contain testing-related keywords (e.g., "test" and "testing") as test files. Non-code files, such as txt or json files, are excluded from this criterion.

After applying deduplication and the above three filtering criteria, 485 vulnerability records out of 37,284 vulnerability records that have more than one commit are retained. Among these, 216 come from PyPI, 204 from Maven, and 65 from npm. This sharp reduction highlights the scarcity of high-quality backports.
Next, we consider all combinations of valid commits in the same vulnerability record as potential backporting task instances. 
This process produces 1,527 unique commit pairs, which will be validated in the following section to determine whether they truly represent backporting relationships.

\subsection{Validating Backporting Data}
\label{sec:validating}

Even after filtering, two commits in the same repository do not necessarily mean they have a backporting relationship. For example, they could be sequential patches for the same vulnerability or completely unrelated commits. To address this, two of our authors independently reviewed all commit pairs, and then compared and discussed any differences in our assessments.
Specifically, we examine the commit messages and the code content to determine whether they address the same vulnerability across different versions of the codebase.
Some commit pairs represent a sequence of continuous patches for the same vulnerability within the same version, and we exclude these cases.
We also exclude commits that contain multiple patches for different vulnerabilities or that are mixed with new features.
One thing to note is that, since these commit pairs are initially produced through combinations, some of them may result in meaningless patch pairs.
For example, suppose a vulnerability record references three patches, $P_a$, $P_b$, $P_c$, that are retained after all the filtering steps described above. From the commit messages and semantic versioning information, we learn that $P_a$ is the original patch, while $P_b$ and $P_c$ are two backported patches derived from $P_a$. These three patches form three possible pairs: $(P_a, P_b)$, $(P_a, P_c)$, and $(P_b, P_c)$. Only the first two pairs represent actual backport relationships, while the last one does not.
When multiple valid patches are associated with the same vulnerability record, we identify the original patch by examining the commit message. If the maintainer explicitly states that a commit is backported from another, we use that information. Otherwise, we compare the semantic versions of the commits and treat the one with the latest version as the original. 
Once the original patch is determined, we exclude any patch pairs that do not involve it, such as $(P_b, P_c)$ in the example above, to avoid duplication.
Following this process, 619 commit pairs are confirmed to represent backport relationships.

\subsection{Categorizing Backporting Data}
\label{sec:categorize}
During the manual inspection in Section~\ref{sec:validating}, we additionally categorize the 619 backport commit pairs based on how much the backported patch differs from its original patch. This provides more insights into how high-quality patch backporting instances are distributed in real-world software development (as further discussed in Section~\ref{sec:rq1}).
Since our filtering ensures that each retained backported patch modifies at least one non-testing file, we perform the categorization at two levels: first at the file level, and then at the content level.
At the file level, we focus on how the modified non-test files in the backported patch map to those in the original patches. Based on this comparison, we define five categories that capture differences at the file level.
\begin{itemize}[leftmargin=*]
    \item \textbf{Files matched (91.9\%):} The files modified in the original patch and the backported patch can be matched accordingly;
    \item \textbf{Backport adds files (2.4\%):} The backported patch modifies additional files not touched by the original patch;
    \item \textbf{Backport omits files (3.9\%):} Some files modified in the original patch are not modified in the backported patch; 
    \item \textbf{Backport adds and omits files (1.3\%):} The backported patch modifies some additional files, while omitting some files that were modified in the original patch.
    \item \textbf{Completely different files (0.5\%):} The set of modified files in the original patch and the backported patch are completely different.
\end{itemize}
For categorizing content-level differences in non-test files, we also define five categories, which are similar to prior work~\cite{shariffdeen2021automated}.

\begin{figure*}[t]
    \centering
    \includegraphics[width=1\linewidth]{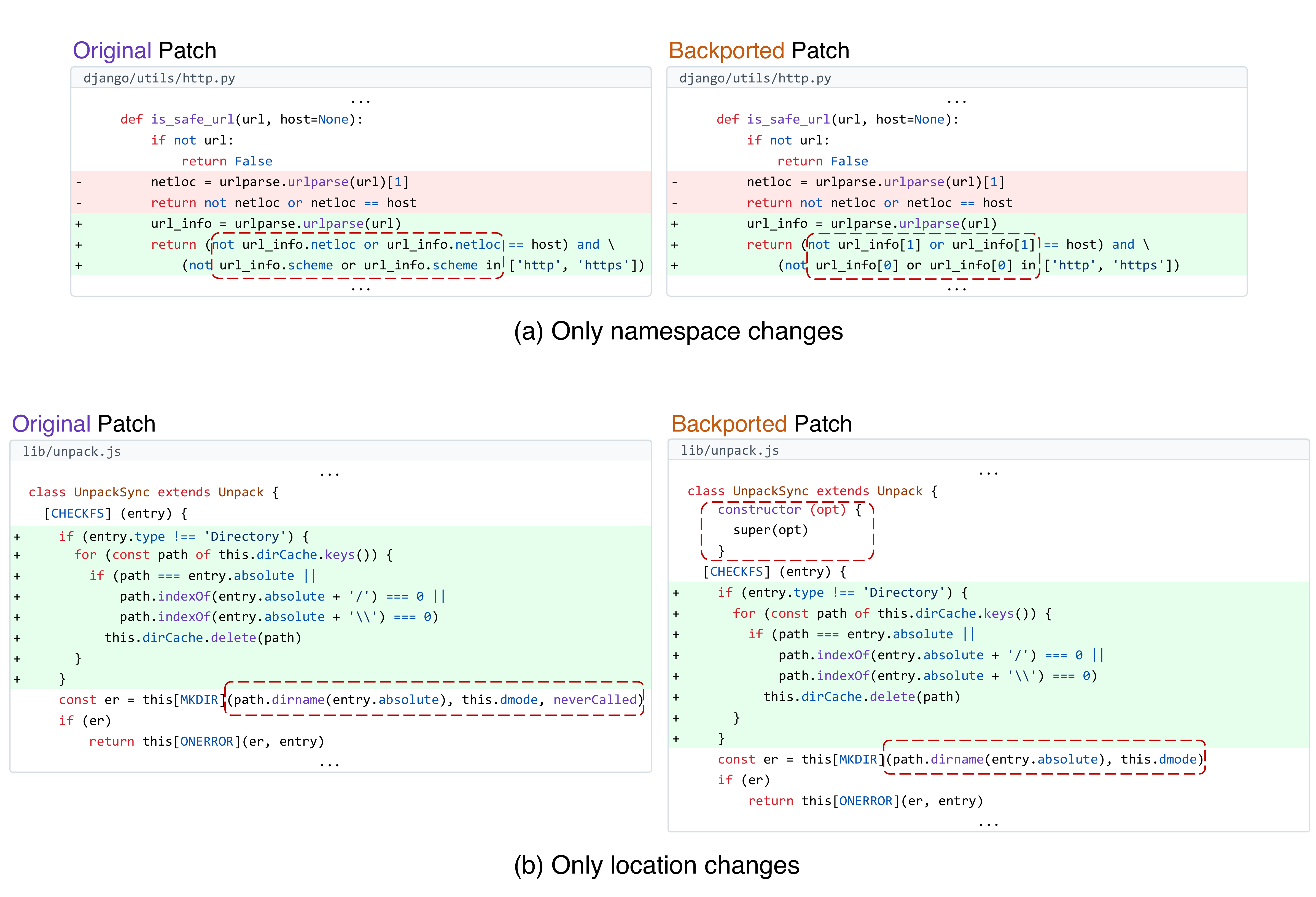}
    \caption{Showcase of only namespace changes and only location changes.}
    \label{fig:category_a_b}
\end{figure*}

\begin{figure*}[t]
    \centering
    \includegraphics[width=1\linewidth]{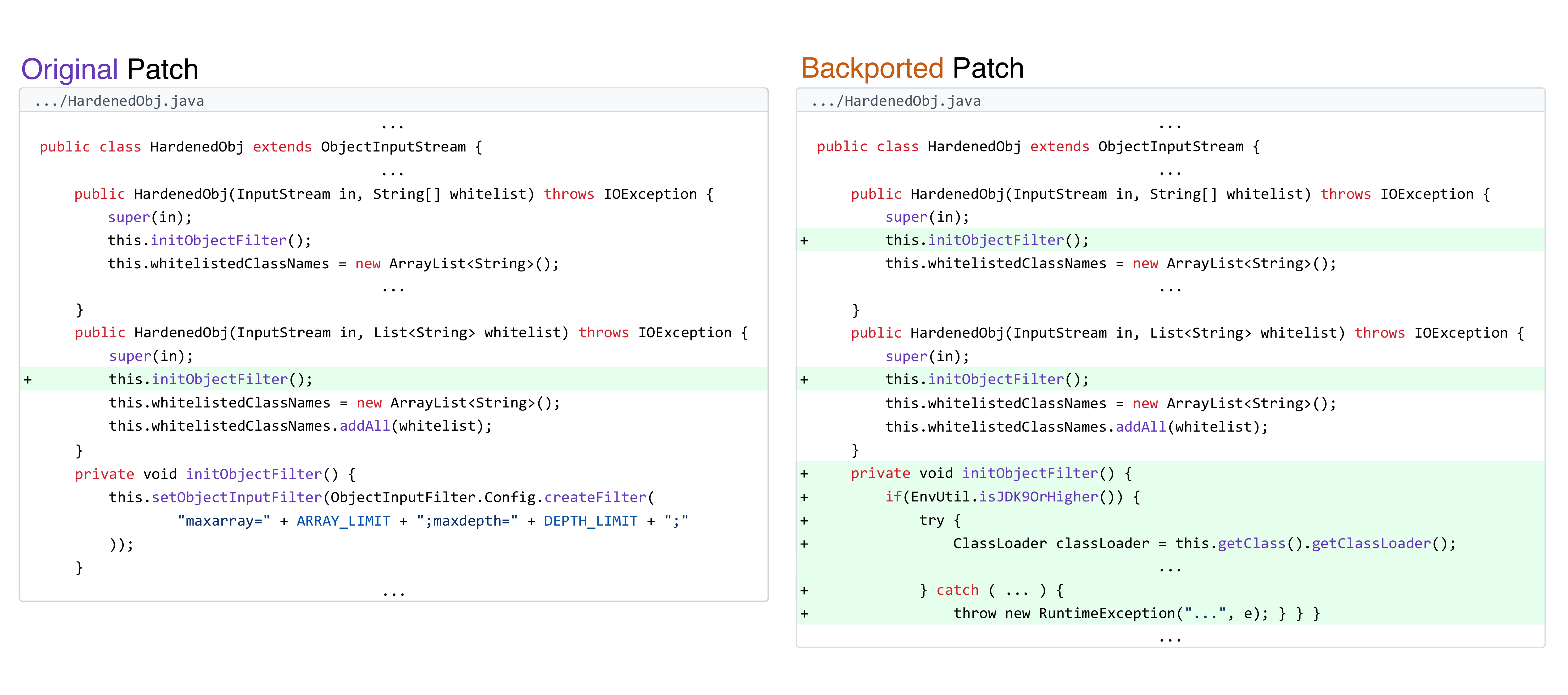}
    \caption{Showcase of logical or structural changes.}
    \label{fig:category_c}
\end{figure*}

\begin{itemize}[leftmargin=*]
    \item \textbf{No changes (48.0\%):} The content of the backported patch and its surrounding code context in the files remain unchanged. In this paper, we define code context from the perspective of the Abstract Syntax Tree (AST) concept. If we parse a program into an AST, each code statement becomes a statement node under a function node, a class node, or a program node. For a backported patch, we define its code context as all sibling nodes that share the same parent node, whether that parent is a function, class, or program node. 
    \item \textbf{Only patch location changes (25.8\%):} The files or code context related to the backported patch have changed, but the patch content itself remains identical to the original patch. Figure~\ref{fig:category_a_b}(b) illustrates a case where the backported patch content remains unchanged, while the code context of the patch has changed, requiring the patch to be relocated. 
    \item \textbf{Only namespace changes (1.0\%):} The backported patch is adapted from the original patch by renaming identifiers such as variables or functions. Figure~\ref{fig:category_a_b}(a) shows an example where the only required modification for the backported patch is related to the namespace. Specifically, the object \texttt{url\_info} changes from a class object to a subscriptable object in the backported codebase, requiring adjustments to how its values are accessed.
    \item \textbf{Location \& namespace changes (3.4\%):} The backported patch is adapted from the original patch by modifying both its location and its namespace.
    \item \textbf{Logical or structural changes (21.8\%):} The backported patch is adapted from the original patch by introducing logical or structural changes. Figure~\ref{fig:category_c} shows an example where the original patch fixes a vulnerability by simply adding a missing filter initialization call to a function. In contrast, the backported patch must first implement the filter initialization function, since it does not exist in the backported codebase, and its implementation differs from that in the original codebase. In addition, the backported patch adds an extra filter initialization call to another function, while this call already exists in the original codebase.
\end{itemize}



\subsection{Execution-Based Validation}
\label{sec:building_env}
Next, we validate the usability of each backporting instance by running its test suite. This process is broken down into three steps. 
First, to balance manual effort with benchmark size, we select four repositories from each ecosystem and validate all backporting instances within them.
Second, we check whether the backported patch can be successfully applied and executed on the backported codebase.
Finally, we analyze the execution log of the related test suite for each instance to confirm its usability for evaluation.

For repository selection, we choose a subset rather than all repositories for validation, since building a virtual environment for each backported codebase requires substantial effort. 
We observed that the setup process is often similar across different versions of the same repository. Based on this observation, we first rank all repositories by the number of confirmed backporting instances and then select the top four from each ecosystem. 
During setup, however, we fail to set up environments for all backporting instances in a few repositories. In such cases, we exclude those repositories and instead select the next repositories with the highest number of confirmed backporting instances in the corresponding ecosystem.
In total, the backporting instances within the twelve repositories we selected account for 51.1\% of all confirmed backporting instances.

In the second step, we roll back the codebase to the parent commit of the backported commit. We then set up the environment following the installation instructions for that version of the codebase. 
Similar to SWE-bench~\cite{jimenez2023swe}, we split the backported patch into a test patch and a gold patch. The test patch contains all modifications to the test files, while the gold patch contains the remaining modifications.
We also write test scripts based on the repository's testing instructions. These scripts run all test cases in the files modified by the test patch.
To ensure the usability of the test cases, we apply the test patch and execute the test scripts before and after the gold patch is subsequently applied, obtaining the log results $log_{pre}$ and $log_{post}$ separately.
If the test scripts run successfully and produce valid test results both before and after the gold patch is applied, we consider the repository to be successfully set up. We discard any backporting instance where all test cases fail both before and after the gold patch is applied, as this suggests that the setup process has failed.

In the third step, we parse $log_{pre}$ and $log_{post}$ to extract the status of each test case. Because the logging format of backporting instances within the same repository is usually consistent, we define one or several regular expression-based log parsers for each repository. These parsers parse the logs and extract the outcome of each test case as either \texttt{fail} or  \texttt{pass}. Noting that some test cases are marked as \texttt{fail} because they are executed in either before or after applying the gold patch, but are skipped or missing in the other.
We consider a test suite usable if at least one test case changes from \texttt{fail} before the gold patch to \texttt{pass} after it. 
Backporting instances without such test cases are excluded from our evaluation dataset.
Following prior work~\cite{jimenez2023swe}, we also exclude backporting instances where failures before the gold patch are due to \texttt{ImportError} and \texttt{AttributeError} in Python, or \texttt{Cannot find symbol} errors in Java. These errors are typically related to dependency naming issues, which are trivial and nearly impossible to resolve automatically.
After the above validation, we record two lists of test cases as evaluation assets for each backporting instance.
The first list, \texttt{FAIL\_TO\_PASS}, contains test cases that change from \texttt{fail} to \texttt{pass}. These test cases serve to verify whether the issue in the backporting task instance is solved by the patch.
The second list, \texttt{PASS\_TO\_PASS}, contains test cases that remain \texttt{pass}, ensuring that the original functionality of related modules is properly maintained.
During evaluation, a backport task is considered resolved if all tests in \texttt{FAIL\_TO\_PASS} and \texttt{PASS\_TO\_PASS} pass.

Overall, we built BackportBench with 202 backporting task instances, where the environments were successfully set up and the usability of their test cases was validated through execution.
We present the final breakdown of these validated backporting instances in Figure~\ref{fig:task-distribution}. We also present related attributes of BackportBench instances from each software ecosystem in Table~\ref{table:dataset-stats}.

\begin{table}[h]
\begin{minipage}{.40\linewidth}
    \begin{minipage}{\textwidth}
        \centering
        \includegraphics[width=\textwidth]{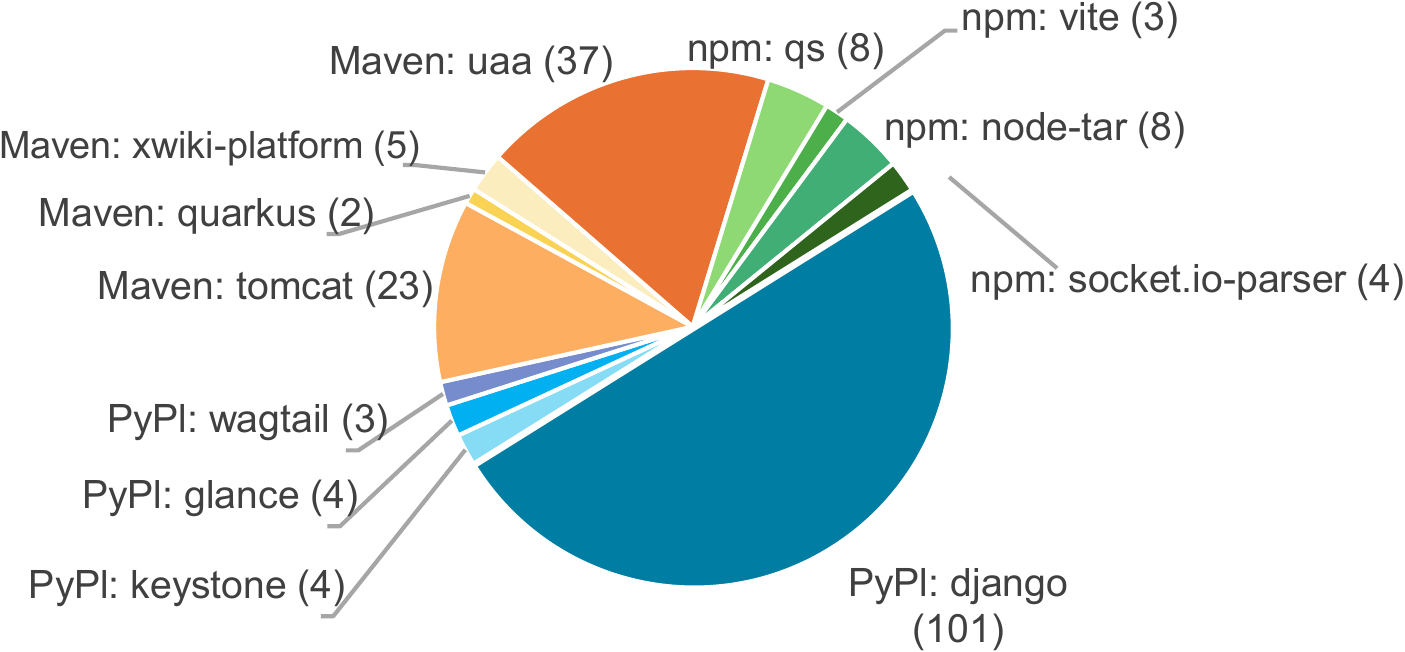}
        \captionof{figure}{
        Distribution of BackportBench tasks from 12 open source packages across the PyPI, Maven, and npm ecosystems.
        }
        \label{fig:task-distribution}
    \end{minipage}
\end{minipage}
\hfill
\begin{minipage}{.58\linewidth}
    \begin{minipage}{\textwidth}
        \centering
        \caption{
        Average and maximum numbers characterizing different attributes of a BackportBench task instances.
        }
        \resizebox{\textwidth}{!}{
        \small
        \begin{tabular}{llrr|rr|rr}
        \toprule
         & & \multicolumn{2}{c}{\textbf{PyPI}}&\multicolumn{2}{c}{\textbf{Maven}} & \multicolumn{2}{c}{\textbf{npm}}\\
            & & Mean & Max & Mean & Max & Mean & Max\\
        \midrule
            \multirow{2}{*}{Codebase} & \# Files (non-test) & 3,538 & 4,330 & 2,922 & 10,000 & 300 & 1,716 \\
                                      & \# Lines (non-test) & 1,666K  & 1,891K & 1,218K  & 4,736K & 197K  & 451K \\
            \midrule
             Original & \# Files edited & 5.8 & 15 & 4.8 & 24 & 2.8 & 6 \\
            Gold Patch                            & \# Lines edited & 54.8 & 174  & 47.7 & 217  & 19.5 & 86 \\
                                       
            \midrule
            Backported & \# Files edited & 4.9 & 15 & 4.9 & 24 & 2.8 & 6 \\
            Gold Patch                            & \# Lines edited & 42.4 & 174  & 50.6 & 217  & 21.4 & 98 \\
                                       
            \midrule
            \multirow{2}{*}{Tests} & \# Fail to Pass & 2.2 & 11 & 2.1 & 10 & 3.2 & 6 \\
                                   & \# Total & 57.6 & 735  & 297.7 & 1,365  & 73.4 & 219 \\
        \bottomrule
        \end{tabular}
        }
        \label{table:dataset-stats}
    \end{minipage}
\end{minipage}
\vspace{-0.5em}
\end{table}

\section{Evaluation}
In this section, we focus on addressing the following three research questions:
\begin{itemize}[leftmargin=*]
    \item \textbf{RQ1:} How are high-quality patch backports distributed in real-world software development?
    \item \textbf{RQ2:} How do LLMs perform on BackportBench?
    \item \textbf{RQ3:} How do existing patch porting methods perform on BackportBench?

\end{itemize}

\subsection{How are high-quality backports distributed in real-world software development?}
\label{sec:rq1}

Since patch backporting, especially those high-quality backports is rare, this RQ examines the distribution of backports to understand when developers adopt this practice.
First, we measure the distribution and severity of vulnerabilities that contain high-quality backports.
Then we analyze the relationship between backported patches and their original patches.

We found that 30.1\% of vulnerability records are in the npm ecosystems, but npm accounts for only 11.5\% of the 619 high-quality backports across the three ecosystems. By contrast, PyPI has fewer reported vulnerabilities (27.7\% of the total) but contains nearly four times as many high-quality backports as npm (42.7\%). Maven account for the remaining 45.9\% of high-quality backports.
Vulnerabilities that report malicious packages were excluded from these results, since such cases typically suggest package removal rather than patching~\cite{osv2025openssf}.
For severity, we used the CVSS scores~\cite{nvd2025vulnerability} recorded in the OSV vulnerability records from which the backported patches were collected.
We found that nearly half (47.0\%) of these backported patches address vulnerabilities rated Critical or High severity, which is similar to the proportion of Critical and High severity among all vulnerabilities across the three ecosystems (52.4\%).

Regarding the relationship between the backported patches and their original patches, we first categorize and analyze differences in modified files and contents (Section~\ref{sec:categorize}). Based on that categorization, we then examine the time elapsed between the initial releases of the original and backported patches and the version gap between them.
In Section~\ref{sec:categorize}, we performed the categorization at both file and content levels. At the file-level, 91.9\% high-quality backports modify the same files as the original patches.
Because most high-quality backports fall into this category, we provide a breakdown of this type in Table~\ref{tab:gaps} for further analysis.
We also parse the first Semantic Versioning (SemVer) commit tags that include the patches with a SemVer parser.
A SemVer number follows the form \texttt{MAJOR.MINOR.PATCH}. Versions with different \texttt{MAJOR} number indicates incompatible API changes between them, while versions with different \texttt{MINOR} numbers indicate added functionality that remains backward compatible within the same major version train~\cite{SemVer}.
To calculate the time duration between backported patches and their original patches, we extract the commit date.

\begin{table}[ht]
    \centering
    \begin{tabularx}{\linewidth}{@{}c 
            >{\centering\arraybackslash}X 
            >{\centering\arraybackslash}X 
            >{\centering\arraybackslash}X 
            >{\centering\arraybackslash}X 
            >{\centering\arraybackslash}X @{}}
        \toprule
        & \thead{No\\Changes}
        & \thead{Only Patch\\Location}
        & \thead{Only\\Namespace}
        & \thead{Location \&\\Namespace}
        & \thead{Logical or\\Structural} \\
        \midrule
        \texttt{Total} & 297 & 160 &6 & 18 & 88\\
        \texttt{Cross Major} & 142 (47.8\%) & 87 (54.4\%) & 4 (66.7\%) &8 (44.4\%) & 51 (58.0\%)\\
        \texttt{Cross Minor} & 152 (51.2\%) & 73 (45.6\%) &2 (33.3\%) &10 (55.6\%) & 37 (42.0\%) \\
        \midrule
        \texttt{Major Gap} & 0.6 / 1.0 & 0.7 / 1.0 &0.8 / 1.2 &0.5 / 1.0 & 0.8 / 1.0 \\
        \texttt{Duration (days)} & 3.5 / 0.0 & 6.3 / 1.2 & 0.0 / 0.0 & 2.3 / 0.8 & 10.2 / 4.0  \\
        \bottomrule
    \end{tabularx}
    \caption{Distribution and statistics for 569 high-quality backports whose modified files match the original patches. For each category, we report the number and percentage of backports that were released in a different major version than the original patch, or in a different minor version within the same major version as the original patch. The major-version gap and the release duration are summarized by the mean and the 75th percentile. Three backports that made no changes were released in the same major and minor version as their original patches.}
    \label{tab:gaps}
\end{table}
Table~\ref{tab:gaps} reveals that 95.8\% of file-matched backports fall into three content-level categories, in descendingly order by frequency, no-change backports, location-change backports, and logical/structural-change backports. This ordering suggests a decreasing willingness among developers to perform larger changes when backporting.
For these three categories, the proportion of backports that cross major versions increases with the extent the changes, while the proportion that only crosses minor versions within the same major version decreases. This is expected, since larger changes typically reflect greater differences between codebase versions and therefore more likely to introduce incompatibilities.
A similar upward trend is observed in the time required for these three categories.
Notably, changing only the patch location (without altering content) indeed increases the proportion of backports that cross major versions and the time required.
The upward trend is not observed for backports that involve namespace changes, possibly because these cases are rare (4.2\%) and lack statistical power. This scarcity aligns with the observation from previous work~\cite{shariffdeen2021automated}
Regarding major-version gaps, most high-quality backports either stay within the same major versions as their originals or cross into just one additional major version.
This finding aligns the nature that maintaining two major-version trains usually captures more than 90\% of dependents, founded by a previous study on packages backporting in popular OSS ecosystems~\cite{decan2021back}.


\begin{center}
    \begin{myboxc} \textbf{RQ1 Summary:} Most high-quality backports come from packages in the PyPI and Maven ecosystems. Patch backporting is less prevalent in the npm ecosystem. 47.0\% of the backports address vulnerabilities rated Critical or High severity. As the extent of changes in a backport increases, the number of such backports typically decreases, while the time required and the proportion that cross major versions increase. Most high-quality backports cross into at most one additional major version compared with their originals. 
    \end{myboxc}
\end{center}

\subsection{How do LLMs perform on BackportBench?}
\label{sec:agent_eval}
\textbf{Setup.}
With the emergence of LLM techniques on coding tasks, particularly resolving GitHub issues~\cite{yang2024swe,xia2024agentless,jimenez2023swe}, which involves understanding a repository and requirements and then generating a commit, we are interested in exploring their potential for patch backporting as well.
Following previous studies~\cite{yang2025survey,dong2025survey}, current issue resolution approaches could be divided into two categories: agentic methods and procedural methods.
We evaluate one representative method from each category.
For the agentic method, we select SWE-agent~\cite{yang2024swe}, a widely used open-source agent framework ranked 4th among frameworks whose performance has been checked on SWE-bench Verified~\cite{openai2024swe}. SWE-agent equips an autonomous LLM agent with tools to interact with the codebase environment. In our evaluation, we provide both the original codebase and patch and the backported codebase in the environment, so that SWE-agent can compare them and determine the adaptations needed to backport the patch.
For the procedural approach, we select Agentless~\cite{xia2024agentless}. Agentless first performs hierarchical localization to identify the relevant files, classes, or functions, and specific edit locations, then generates multiple patches and uses synthesized tests to select the best one.
To tailor Agentless to BackportBench tasks, we add the original patches and information from the original codebase (obtained with Agentless's tools for localization) to the prompt.
Since Agentless and SWE-agent were originally implemented for Python, we use MagentLess and MSWE-agent, the Java and JavaScript extensions of Agentless and SWE-agent from Multi-SWE-bench~\cite{zan2025multi}, for Java and JavaScript backporting tasks. 
We apply the same prompt adaptations when evaluating MSWE-agent and Magentless on BackportBench instances. In this paper, we use the label (M)SWE-agent to refer to SWE-agent when evaluating Python instances and MSWE-agent when evaluating Java and JavaScript instances. Likewise, (M)Agentless denotes Agentless for Python instances and Magentless for Java and JavaScript instances.
In addition to (M)SWE-agent and (M)Agentless, we include an Oracle Retrieval method in our evaluation, which is similar to the previous evaluation on repository-level coding tasks studies~\cite{jimenez2023swe,guo2025omnigirl}. 
This method directly supplies the ground-truth modified files and the original patch as input, simulating a scenario where developers accurately know which files to edit. Including it in the evaluation helps us measure how file localization performance affects repository-level coding tasks.
Including this method in the evaluation can help us demystify the impact of file localization accuracy on repository-level coding tasks.

\textbf{Model Selection.} We evaluate three popular LLMs across these three approaches: GPT-5 (GPT-5-Chat 2025-09-07)~\cite{openai2025gp5}, Claude Sonnet 4 (Claude-Sonnet-4-20250514)~\cite{anthropic2025sonnet4}, and Qwen3-Coder (Qwen3-Coder-480B-A35B-Instruct)~\cite{qwen3technicalreport}. These are top-tier models on the SWE-bench Bash Only setting with strong coding capability and long text understanding abilities~\cite{jimenez2023swe}.

\textbf{Metrics.} As mentioned in Section~\ref{sec:building_env}, a backporting task is considered resolved if all test cases in \texttt{FAIL\_TO\_PASS} (F2P) and \texttt{PASS\_TO\_PASS} (P2P) pass. Based on the evaluation outcome, we further break down the failure cases into five types. 
\begin{itemize}[leftmargin=*]
    \item \textbf{Generation Failed.} This usually occurs because the limited capability of baselines to address the difficult cases and fail in half without a valid output.
    \item \textbf{Only F2P Failed.} The generated backported patches fail the F2P test cases but pass the P2P test cases.
    \item \textbf{Only P2P Failed.} The generated backported patches fail the P2P test cases but pass the F2P test cases.
    \item \textbf{Both Failed.} The generated backported patches fail on both F2P tests and P2P tests.
    \item \textbf{Timeout.} The generated backported patches fail due to exceeding twice the maximum duration of a single instance run in the Execution-based Validation step of the construction pipeline shown in Figure~\ref{fig:workflow}.
\end{itemize}

\textbf{Results.} We analyze the resolve rates of issue-resolution techniques in BackportBench from three perspectives: programming language, content-level category, and file-level category. Finally, we examine the failure types of the backported patch they generated.

\begin{table}[htbp]
  \centering
  \small
  \setlength{\tabcolsep}{4pt}
    \caption{Resolve rates for GitHub issue-resolution techniques evaluated on BackportBench across different programming languages. In this table, "GPT-5-Chat-Latest (2025-09-07)" is abbreviated as GPT-5, "Claude-Sonnet-4-20250514" as Claude Sonnet 4, and "Qwen3-Coder-480B-A35B-Instruct" as Qwen3-Coder. Baselines with the highest resolve rate in each programming language are bolded.}
  \label{table:rq3_across_ecosystem}
  \tabmargin
    \begin{tabular}{llcccc}
    \toprule
    & & \thead{Python} & \thead{Java} & \thead{JavaScript} & \thead{Total}  \\
    \cmidrule(r){3-6}
    \multirow{3}{*}{(M)SWE-agent} 
    & GPT-5 & 78.6\% (88/112) & 44.8\% (30/67) & 34.8\% (8/23) &62.4\% (126/202) \\
    & Claude Sonnet 4 & \textbf{91.1\% (102/112)} & 46.3\% (31/67)  & 43.5\% (10/23) & \textbf{70.8\% (143/202)} \\
    & Qwen3-Coder & 82.1\% (92/112) & 43.3\% (29/67) & 56.5\% (13/23) & 66.3\% (134/202)\\
    \midrule
    \multirow{3}{*}{(M)Agentless} 
    & GPT-5 & 68.8\% (77/112)  & 41.8\% (28/67)  & 30.4\% (7/23) & 55.4\% (112/202) \\
    & Claude Sonnet 4 & 56.3\% (63/112)  & 41.8\% (28/67)  & 13.0\% (3/23)  & 46.5\% (94/202) \\
    & Qwen3-Coder & 66.1\% (74/112)  & 40.3\% (27/67)  & 13.0\% (3/23) & 51.5\% (104/202) \\
    \midrule
    \multirow{3}{*}{Oracle Retrieval} 
    & GPT-5 & 75.9\% (85/112)  & \textbf{62.7\% (42/67)}  & 60.9\% (14/23) & 69.8\% (141/202)\\
    & Claude Sonnet 4  & 79.5\% (89/112)  & 53.7\% (36/67)  & \textbf{60.9\% (14/23)} & 68.8\% (139/202)\\
    & Qwen3-Coder & 77.7\% (87/112)  & 56.7\% (38/67)  & 60.9\% (14/23) & 68.8\% (139/202) \\
    \bottomrule
    \end{tabular}
\end{table}
Specifically, Table~\ref{table:rq3_across_ecosystem} shows the performance of three baselines on Python, Java, and JavaScript task instances.
Among the three baselines, (M)SWE-agent with Claude Sonnet 4 achieves the highest resolve rate on the entire BackportBench (70.8\%). This is followed by Oracle Retrieval method with GPT-5 Chat, which achieves 69.8\% resolve rate. (M)Agentless with GPT-5 Chat resolves 55.4\% of the BackportBench task instances. Compared with (M)Agentless, the Oracle Retrieval method shows certain improvements in resolving backporting tasks. This indicates that localization is one of the bottlenecks for (M)Agentless, especially for JavaScript instances.

However, all baselines perform significantly worse on Java instances and JavaScript instances than on Python instances. (M)SWE-agent with Claude Sonnet 4 achieves a high resolve rate of 91.1\% on Python instances, but falls sharply to 46.3\% on Java and 43.5\% on JavaScript instances.
This aligns with the findings in Multi-SWE-bench~\cite{zan2025multi}, suggesting that the current LLMs and issue-resolution methods struggle to generalize beyond Python, potentially because of differences in language complexity and the characteristics of their task instances.

Table~\ref{table:rq3_across_content} shows the breakdown of the resolved instances for each baseline by content-level category.
The results show that (M)Agentless has a decreasing resolve rate for instances that require no changes, only location changes, and logical or structural changes.
Meanwhile, Oracle Retrieval maintains a similar resolve rate for instances that need no changes and those that require only patch-location changes.
\begin{table}[t]
  \centering
  \setlength{\tabcolsep}{4pt}
    \caption{Performance comparison across different \textbf{content-level} categories on BackportBench. In this table, "GPT-5-Chat-Latest (2025-09-07)" is abbreviated as GPT-5, "Claude-Sonnet-4-20250514" as Claude Sonnet 4, and "Qwen3-Coder-480B-A35B-Instruct" as Qwen3-Coder. }
  \label{table:rq3_across_content}
  \tabmargin
  \resizebox{\columnwidth}{!}{
    \begin{tabular}{llccccc}
    \toprule
    & &   \thead{No\\Changes}
        & \thead{Only Patch\\Location}
        & \thead{Only\\Namespace}
        & \thead{Location \&\\Namespace}
        & \thead{Logical or\\Structural}  \\
    \cmidrule(r){3-7}
    \multirow{3}{*}{(M)SWE-agent} 
    & GPT-5 & 61.5\% (64/104) & 75.9\% (41/54) & 66.7\% (2/3) & 66.7\% (6/9) & 40.6\% (13/32)  \\
    & Claude Sonnet 4 & 69.2\% (72/104)  & 75.9\% (41/54)  & 33.3\% (1/3) & 100.0\% (9/9) & 62.5\% (20/32)\\
    & Qwen3-Coder & 66.3\% (69/104) & 77.8\% (42/54) & 33.3\% (1/3) & 88.9\% (8/9) & 43.8\% (14/32)  \\
    \midrule
    \multirow{3}{*}{(M)Agentless} 
    & GPT-5 & 68.3\% (71/104)  & 53.7\% (29/54)  & 33.3\% (1/3) & 44.4\% (4/9) & 21.9\% (7/32)  \\
    
    & Claude Sonnet 4 & 55.8\% (58/104)  & 40.7\% (22/54)  & 66.7\% (2/3) & 66.7\% (6/9) & 18.8\% (6/32)  \\
    
    & Qwen3-Coder & 58.7\% (61/104)  & 50.0\% (27/54)  & 100.0\% (3/3) & 66.7\% (6/9) & 21.9\% (7/32) \\
    \midrule
    \multirow{3}{*}{Oracle Retrieval} 
    & GPT-5 & 76.9\% (80/104)  & 75.9\% (41/54)  & 100.0\% (3/3) & 66.7\% (6/9) & 34.4\% (11/32) \\
    & Claude Sonnet 4  & 73.1\% (76/104)  & 79.6\% (43/54)  & 66.7\% (2/3) & 55.6\% (5/9) & 40.6\% (13/32)  \\
    & Qwen3-Coder & 76.0\% (79/104)  & 75.9\% (41/54)  & 100.0\% (3/3) & 88.9\% (8/9) & 25.0\% (8/32)  \\
    \bottomrule
    \end{tabular}
  }

\end{table}
Surprisingly, (M)SWE-agent techniques resolve instances that require changing the patch location even better than those that need no changes. While RQ1 found that modifying only the patch location increases the time required for human developers, indicating greater effort compared with backports that require no changes. 
In addition, all baselines perform relatively poorly on instances that require logical or structural adaptations. However, (M)SWE-agent still achieves a resolve rate of 40.6\% to 62.5\%, which is the highest among the baselines and two to three times higher than (M)Agentless. This suggests that agentic methods like SWE-agent, which can flexibly plan and execute, are better suited to resolving difficult backporting tasks.

Except for file-matched task instances, BackportBench only contains one instance per file-level category due to their scarcity. The performance of each baseline on each file-level category (except for file-matched instances) is reported in Table~\ref{table:rq3_across_file}. We found that all methods perform well when the files modified in the original patch include the files that need to be modified in the backported patch. In addition, (M)SWE-agent can sometimes resolve cases where the modified files overlap between the original patch and the backported patches.

Finally, we examine the failure types for all issue-resolution approaches on BackportBench. Table~\ref{table:agent_failure_type} shows that the majority of failures are due to failing F2P or P2P tests, indicating the generated backported patches are ineffective. 
However, (M)Agentless and Oracle Retrieval failed to generate a valid patch in 23.0\%-41.7\% of cases. This is mainly because these methods fail to complete the procedure or to match the formatting in the procedure, and they lack a dynamic reflect-and-retry mechanism.



\begin{center}
    \begin{myboxc} \textbf{RQ2 Summary:} Existing issue-resolution methods successfully resolve 70.8\% of backport instances, but performance varies significantly across programming languages. All methods generally perform worse as the extent of required change increases, while (M)SWE-agent performs better on instances that require changing the patch location than on those that need no changes. The majority of the failures are caused by failing test cases, and (M)Agentless and Oracle Retrieval methods fail to generate valid patches in a notable share of instances.
    \end{myboxc}
\end{center}
\begin{table}[t]
  \centering
  \setlength{\tabcolsep}{4pt}
    \caption{Performance comparison across different \textbf{File-level} categories on BackportBench. In this table, "GPT-5-Chat-Latest (2025-09-07)" is abbreviated as GPT-5, "Claude-Sonnet-4-20250514" as Claude Sonnet 4, and "Qwen3-Coder-480B-A35B-Instruct" as Qwen3-Coder. }
  \label{table:rq3_across_file}
  \tabmargin
    \begin{tabular}{llcccc}
    \toprule
    & &   \thead{Files\\Added}
        & \thead{Files\\Omitted}
        & \thead{Added \&\\Omitted}
        & \thead{No\\Overlap}  \\
    \cmidrule(r){3-6}
    \multirow{3}{*}{(M)SWE-agent} 
    & GPT-5 & $\checkmark$ & $\checkmark$ & $\times$ & $\times$  \\
    & Claude Sonnet 4 & $\checkmark$  & $\checkmark$  & $\times$ & $\times$  \\
    & Qwen3-Coder & $\times$ & $\checkmark$ & $\checkmark$ & $\times$   \\
    \midrule
    \multirow{3}{*}{(M)Agentless} 
    & GPT-5 & $\times$ & $\checkmark$ & $\times$ & $\times$  \\
    & Claude Sonnet 4 & $\times$  & $\times$  & $\times$ & $\times$  \\
    & Qwen3-Coder & $\times$ & $\checkmark$ & $\times$ & $\times$   \\
    \midrule
    \multirow{3}{*}{Oracle Retrieval} 
    & GPT-5 & $\times$ & $\checkmark$ & $\times$ & $\times$  \\
    & Claude Sonnet 4 & $\checkmark$  & $\checkmark$  & $\checkmark$ & $\times$  \\
    & Qwen3-Coder & $\times$ & $\checkmark$ & $\times$ & $\times$   \\
    \bottomrule
    \end{tabular}

\end{table}

\begin{table}[t]
  \centering
  \setlength{\tabcolsep}{4pt}
    \caption{The failure types of issue resolution techniques on BackportBench.  In this table, "GPT-5-Chat-Latest (2025-09-07)" is abbreviated as GPT-5, "Claude-Sonnet-4-20250514" as Claude Sonnet 4, and "Qwen3-Coder-480B-A35B-Instruct" as Qwen3-Coder. Failure types with highest frequency for each baseline are bolded.}
  \label{table:agent_failure_type}
  \tabmargin
  \resizebox{\columnwidth}{!}{
    \begin{tabular}{llccccc}
    \toprule
    & & \thead{Generation\\Failed} & \thead{Only F2P\\Failed} & \thead{Only P2P\\Failed} & \thead{Both\\Failed} & \thead{Timeout}   \\
    \cmidrule(r){3-7}
    \multirow{3}{*}{(M)SWE-agent} 
    & GPT-5 & 13.2\% (10/76) & \textbf{39.5\% (30/76)} & 38.2\% (29/76) & 6.6\% (5/76) & 2.6\% (2/76)  \\
    & Claude Sonnet 4 & 20.3\% (12/59)  & \textbf{64.4\% (38/59)}  & 6.8\% (4/59) & 5.1\% (3/59) & 3.4\% (2/59) \\
    & Qwen3-Coder & 29.4\% (20/68) & \textbf{36.8\% (25/68)} & 22.1\% (15/68) & 7.4\% (5/68) & 4.4\% (3/68) \\
    \midrule
    \multirow{3}{*}{(M)Agentless} 
    & GPT-5 & 28.9\% (26/90)  & \textbf{35.6\% (32/90)}  & 32.2\% (29/90) & 2.2\% (2/90) & 1.1\% (1/90)  \\
    
    & Claude Sonnet 4 & \textbf{41.7\% (45/108)}  & 33.3\% (36/108)  & 18.5\% (20/108) & 5.6\% (6/108) & 0.9\% (1/108)  \\
    & Qwen3-Coder & 34.7\% (34/98)  & \textbf{39.8\% (39/98)}  & 18.4\% (18/98) & 7.1\% (7/98) & 0.0\%(0/98) \\
    \midrule
    \multirow{3}{*}{Oracle Retrieval} 
    & GPT-5 & 23.0\% (14/61)  & \textbf{42.6\% (26/61)}  & 31.1\% (19/61) & 1.6\% (1/61) & 1.6\% (1/61) \\
    & Claude Sonnet 4  & \textbf{33.3\% (21/63)}  & 30.2\% (19/63)  & 30.2\% (19/63) & 4.8\% (3/63) & 1.6\% (1/63)  \\
    & Qwen3-Coder & \textbf{33.3\% (21/63)}  & 31.7\% (20/63)  & 25.4\% (16/63) & 7.9\% (5/63) & 1.6\% (1/63) \\
    \bottomrule
    \end{tabular}
    }

\end{table}





\subsection{How do existing patch porting methods perform on BackportBench?}
\label{sec:trad}

As we mentioned in Section~\ref{sec:tra_baselines}, most of the previous methods focus on porting patches in C. Among which only \mystique and \ppathf are compatible with Java project~\cite{wu2025mystique, pan2024automating}. 
Therefore, we evaluate the performance of these two prior patch porting methods on the Java task instances in BackportBench.
Since these two methods were originally designed for function-level tasks, we extract the related modules that include functions from the original patches and the backported patches following their input format. Then we iteratively generate the backported patch for each function and merge them.

Similar to RQ2, we use the resolve rate on F2P and P2P test cases to evaluate the generated backported patches, and break down the resolved cases by content-level categories. We also analyze the failure types of \mystique and \ppathf.

Table~\ref{table:trad_across_content} shows the performance of \ppathf and \mystique on backport instances across content-level categories and compares them with the MSWE-agent.
The results show that \mystique, the previous state-of-the-art method for patch porting, can resolve 40.3\% of Java task instances, while MSWE-agent with GPT-5 Chat can resolve 46.3\%, a 15\% relative improvement over \mystique. The other baseline, \ppathf can resolve only a single no-change instance.

We further examine the specific categories of instances they resolved. The results show that \mystique resolves 55.8\% of the no-change instances, which is one more case than the best MSWE-agent setting (Claude Sonnet 4). Nevertheless, when the tasks become slightly more complex and require location changes, the resolve rate of \mystique falls significantly to 22.2\%, resolving only 2 of 9 instances. By contrast, MSWE-agent with GPT-5 Chat and with Qwen3-Coder-480B resolve 7 (77.8\%) and 6 (66.7\%) of those instances, respectively. 
For the most difficult cases that require logical or structural changes, \mystique resolves only 1 instance (7.7\%). While MSWE-agent with Claude Sonnet 4 and with GPT-5 Chat can still manage to resolve 5 (38.5\%) and 4 (30.8\%) instances, respectively.
\begin{table}[t]
  \centering
  \setlength{\tabcolsep}{4pt}
    \caption{The resolve rate of exiting patch porting methods compared with MSWE-agent on \textbf{Java instances} across different content-level categories. Noted that there is no Java task instance that belongs to Location \& Namespace changes. In this table, MSWE.GPT5 is short for MSWE-agent with GPT-5-Chat-Latest (2025-09-07), MSWE.Claude4 is short for MSWE-agent with Claude-Sonnet-4-20250514, and MSWE.Qwen3 is short for MSWE-agent with Qwen3-Coder-480B-A35B-Instruct. The highest resolve rate for each category is bolded.}
    
  \label{table:trad_across_content}
  \tabmargin
  \resizebox{\columnwidth}{!}{
    \begin{tabular}{lcccccc}
    \toprule
    &    \thead{No\\Changes}
        & \thead{Only Patch\\Location}
        & \thead{Only\\Namespace}
        & \thead{Location \&\\Namespace}
        & \thead{Logical or\\Structural}
        & \thead{Total}  \\
    \midrule
     \ppathf & 2.3\% (1/43)  & 0.0\% (0/9)  & 0.0\% (0/2) &N/A& 0.0\% (0/13) & 1.5\% (1/67)\\
     \mystique & \textbf{58.1\% (25/43)}  & 22.2\% (2/9)  & 0.0\% (0/2)  &N/A& 7.7\% (1/13) & 41.8\% (28/67)\\
    \midrule
     MSWE.GPT5 & 41.9\% (18/43) & \textbf{77.8\% (7/9)}  & \textbf{50.0\% (1/2)} &N/A& 30.8\% (4/13) & 44.8\% (30/67) \\
     MSWE.Claude4 &53.5\% (23/43)  & 33.3\% (3/9)  & 0.0\% (0/2)  &N/A& \textbf{38.5\% (5/13)} & \textbf{46.3\% (31/67)} \\
     MSWE.Qwen3 & 48.8\% (21/43) & 66.7\% (6/9) & 0.0\% (0/2) & N/A& 15.4\% (2/13) & 43.3\% (29/67) \\

    \bottomrule
    \end{tabular}
  }

\end{table}

\begin{table}[t]
  \centering
  \setlength{\tabcolsep}{4pt}
    \caption{The failure types of existing patch porting methods on \textbf{Java instances}. }
  \label{table:trad_fail_reasons}
  \tabmargin
    \begin{tabular}{lccccc}
    \toprule
    &    \thead{Generation\\Failed}
        & \thead{Only F2P\\Failed}
        & \thead{Only P2P\\Failed}
        & \thead{Both\\Failed}
        & \thead{Timeout}\\
    \cmidrule(r){2-6}
     \ppathf & \textbf{84.8\% (56/66)}  & 10.6\% (7/66)  & 0.0\% (0/66) & 1.5\% (1/66) & 3.0\% (2/66)\\
     \mystique & 12.8\% (5/39)  & 5.1\% (2/39)  & \textbf{74.4\% (29/39)} & 0.0\% (0/39) & 7.7\% (3/39) \\

    \bottomrule
    \end{tabular}

\end{table}
Similar to Section~\ref{sec:agent_eval}, we further look into the failure report in Table~\ref{table:trad_fail_reasons} to find more insights on how \ppathf and \mystique failed.
We found that most (84.8\%) of \ppathf's failures are due to generation failure. Specifically, misaligning the target path or not finding the corresponding function between the original and the backported patch.
But \ppathf often overlooks the diverse reasons for LLM porting failure, which hinders further patch adaptation and leads to unsuccessful patching. Moreover, \ppathf relies on LLM fine-tuning, which is unable to proceed due to the limited number of BackportBench, thereby reducing its effectiveness across the backporting task and the Java language. Those factors ultimately affect the evaluation results of \ppathf.

Interestingly, the failure report for \mystique shows that although it can successfully generate the backported patches for most cases and subsequently pass the F2P test cases, most of its failure are due to failing on P2P test cases. This implies that the backported patches it generates are likely to break the original functionality of the modified classes. However, this negative impact is ignored by all the previously patch porting methods that use equivalence-based or similarity-based as evaluation metric~\cite{shariffdeen2020automated,shariffdeen2021automated,pan2024automating,yang2023enhancing,wu2025mystique}.
We believe that it is because \mystique adopts a function-level backporting design that does not consider the context across the functions within the same file and context across files. 



\begin{center}
    \begin{myboxc} \textbf{RQ3 Summary:} Previous state-of-the-art patch porting method can resolve half of the no-changes Java backporting tasks, but still lack the capability to resolve more complicated cases even only the patch location should modified.
    \end{myboxc}
\end{center}

\section{Discussion}
\subsection{Implications for Researchers}
This benchmark serves as the first comprehensive repository-level, multilingual benchmark that evaluates the effectiveness of patch backporting techniques. In our evaluation, we found that both issue-resolution techniques and previous patch porting techniques perform relatively poorly at tasks that need logical or structural changes, indicating a major challenge that future research on developing patch backporting techniques needs to focus on. Nevertheless, the amount of such task instances in BackportBench is low, since developers typically need to take more time and resources on such cases and relatively have less willingness to perform such backporting. To facilitate the automated patch backporting techniques as reliable techniques in real-world software development, future research on this topic should also try to extend Benchmark with more difficult cases to ensure a more diverse evaluation of these techniques.

Beyond patch backporting, BackportBench can also be used to evaluate repository-level code evolution and migration. Patch backporting, code migration, and code evolution share common automation challenges: identifying changes and incompatibilities between versions, and adapting code to the target version. However, no existing evaluation method for these tasks covers repository-level scenarios and multiple programming languages simultaneously~\cite{liu2025migrationbench,cheng2025codemenv}. 
Future research on repository-level code evolution and migration can integrate BackportBench for more comprehensive evaluation.

From the results of RQ2, we found that the agentic approach (represented by SWE-agent) outperforms the procedural approach on backporting tasks, potentially due to the dynamic reflection-and-retry mechanism. In addition, existing procedural methods do not adequately support LLMs in comparing and understanding how a codebase evolves versions, and subsequently fail to determine the adaptation needed for backporting patches. We suggest future research on automated patch backporting add more support for comparing the differences between different codebase versions.

Regarding the applications of patch backporting techniques, we suggest that researchers further explore whether smoother vulnerability patch propagation to software users can really be achieved with the development of future automated backporting techniques.
As recent research~\cite{wang2023plumber,zhong2025empirical} discussed, whenever a vulnerability is found and fixed in a software package, its downstream users have difficulties upgrading to the patched versions as they are typically versions in different major version trains. Upgrading may need to address incompatibility issues, which may be impossible for unmaintained dependent packages.
Future research on automated backporting techniques can validate whether deploying these generated backported patches to a backward-compatible release can really help the users of vulnerable packages upgrade smoothly, subsequently remove the vulnerability without introducing incompatibilities. 

\subsection{Implication for Software Practitioners}
For software practitioners, the finding in Section~\ref{sec:trad} suggests that test cases for validating the backported patches should be included while performing backporting. As software developers usually backport patches completely based on the original patches, which could introduce inconsistency between the patched code and the context across versions~\cite{xie2024unveiling}. We suggest that the developers should carefully test the backported patches even when the surrounding code context looks similar.

\section{Threats to Validity}
\subsection{Internal Validity}
This paper suffers from several internal threats. First, like previous repository-level coding benchmarks~\cite{zan2025multi,guo2025omnigirl} that rely on manual construction of executable environments, the environments we build may not exactly match those that existed when developers performed backporting. This is mainly due to the loose version constraints on package dependencies and system dependencies, which can cause different versions to be installed over time, and to insufficient documentation and guidelines.
However, we validate the usability of each task instance by running tests, which should mitigate the impact of inconsistent environment reproduction. In addition, all BackportBench task instances are released as Docker containers with all dependencies installed, which facilitates consistent evaluation in future research.
Second, throughout the BackportBench construction, we tried to collect as much realistic backport data as possible. However, due to the scarcity of high-quality backports, many of them were conducted predate the LLM's knowledge cutoff date, which may introduce data contamination into the evaluation results. To mitigate this, we open-source our data collection and curation code to facilitate dynamic updates for BackportBench. Inspired by recent work~\cite{zhang2025swe,guo2025swe}, we plan to develop automated approaches to identify and construct backport instances, further reducing the manual effort required for live updates to BackportBench in the future.

\subsection{External Validity}
As shown in Section~\ref{sec:agent_eval}, agents that use different LLMs can vary in their patch-backporting performance on different programming languages. This variation could limit the generalizability of our findings.
To mitigate this threat, we constructed BackportBench using data collected from the three most popular programming languages~\cite{cass2025top}. Each of these languages has a large software ecosystem with hundreds of thousands of reusable packages, so we believe this selection covers the most common patch-backporting scenarios.
The conclusion about existing approaches' performance may also depend on the packages used for evaluation. This is because each package follows its own maintenance strategy and tends to perform similar types of backports (for example, making no code changes) across a specific range of versions, which can bias the results.
To increase the diversity and obtain sufficient task instances, we construct task instances from 12 repositories spanning three programming languages.

\section{Conclusion}
In this paper, we present BackportBench, the first patch backporting benchmark suite comprising 202 real-world patch backporting problems curated from PyPI, Maven, and npm. BackportBench reframes the evaluation of patch backporting by using execution-based testing. Using BackportBench, we evaluate three widely used GitHub issue-resolution approaches with three state-of-the-art LLMs, along with two existing patch porting methods, and report several notable findings. We also provide an in-depth discussion of implications for both researchers and software practitioners.

\section{Data Availability}
The code and data of BackportBench are available at \url{https://github.com/BackportBench/BackportBench}.

\bibliographystyle{ACM-Reference-Format}
\bibliography{Reference}

@inproceedings{shariffdeen2021automated,
  title={Automated patch backporting in Linux (experience paper)},
  author={Shariffdeen, Ridwan and Gao, Xiang and Duck, Gregory J and Tan, Shin Hwei and Lawall, Julia and Roychoudhury, Abhik},
  booktitle={Proceedings of the 30th ACM SIGSOFT International Symposium on Software Testing and Analysis},
  pages={633--645},
  year={2021}
}

@inproceedings{yang2023enhancing,
  title={Enhancing oss patch backporting with semantics},
  author={Yang, Su and Xiao, Yang and Xu, Zhengzi and Sun, Chengyi and Ji, Chen and Zhang, Yuqing},
  booktitle={Proceedings of the 2023 ACM SIGSAC Conference on Computer and Communications Security},
  pages={2366--2380},
  year={2023}
}

@article{wu2025mystique,
  title={Mystique: Automated Vulnerability Patch Porting with Semantic and Syntactic-Enhanced LLM},
  author={Wu, Susheng and Wang, Ruisi and Cao, Yiheng and Chen, Bihuan and Zhou, Zhuotong and Huang, Yiheng and Zhao, JunPeng and Peng, Xin},
  journal={Proceedings of the ACM on Software Engineering},
  volume={2},
  number={FSE},
  pages={130--152},
  year={2025},
  publisher={ACM New York, NY, USA}
}

@inproceedings{pan2024automating,
  title={Automating zero-shot patch porting for hard forks},
  author={Pan, Shengyi and Wang, You and Liu, Zhongxin and Hu, Xing and Xia, Xin and Li, Shanping},
  booktitle={Proceedings of the 33rd ACM SIGSOFT International Symposium on Software Testing and Analysis},
  pages={363--375},
  year={2024}
}

@inproceedings{bhuiyan2023secbench,
  title={SecBench. js: An executable security benchmark suite for server-side JavaScript},
  author={Bhuiyan, Masudul Hasan Masud and Parthasarathy, Adithya Srinivas and Vasilakis, Nikos and Pradel, Michael and Staicu, Cristian-Alexandru},
  booktitle={2023 IEEE/ACM 45th International Conference on Software Engineering (ICSE)},
  pages={1059--1070},
  year={2023},
  organization={IEEE}
}

@article{jimenez2023swe,
  title={Swe-bench: Can language models resolve real-world github issues?},
  author={Jimenez, Carlos E and Yang, John and Wettig, Alexander and Yao, Shunyu and Pei, Kexin and Press, Ofir and Narasimhan, Karthik},
  journal={arXiv preprint arXiv:2310.06770},
  year={2023}
}

@article{zan2025multi,
  title={Multi-swe-bench: A multilingual benchmark for issue resolving},
  author={Zan, Daoguang and Huang, Zhirong and Liu, Wei and Chen, Hanwu and Zhang, Linhao and Xin, Shulin and Chen, Lu and Liu, Qi and Zhong, Xiaojian and Li, Aoyan and others},
  journal={arXiv preprint arXiv:2504.02605},
  year={2025}
}

@article{guo2025omnigirl,
  title={Omnigirl: A multilingual and multimodal benchmark for github issue resolution},
  author={Guo, Lianghong and Tao, Wei and Jiang, Runhan and Wang, Yanlin and Chen, Jiachi and Liu, Xilin and Ma, Yuchi and Mao, Mingzhi and Zhang, Hongyu and Zheng, Zibin},
  journal={Proceedings of the ACM on Software Engineering},
  volume={2},
  number={ISSTA},
  pages={24--46},
  year={2025},
  publisher={ACM New York, NY, USA}
}

@article{xia2024agentless,
  title={Agentless: Demystifying llm-based software engineering agents},
  author={Xia, Chunqiu Steven and Deng, Yinlin and Dunn, Soren and Zhang, Lingming},
  journal={arXiv preprint arXiv:2407.01489},
  year={2024}
}

@article{yang2024swe,
  title={Swe-agent: Agent-computer interfaces enable automated software engineering},
  author={Yang, John and Jimenez, Carlos E and Wettig, Alexander and Lieret, Kilian and Yao, Shunyu and Narasimhan, Karthik and Press, Ofir},
  journal={Advances in Neural Information Processing Systems},
  volume={37},
  pages={50528--50652},
  year={2024}
}

@article{decan2021back,
  title={Back to the past--analysing backporting practices in package dependency networks},
  author={Decan, Alexandre and Mens, Tom and Zerouali, Ahmed and De Roover, Coen},
  journal={IEEE Transactions on Software Engineering},
  volume={48},
  number={10},
  pages={4087--4099},
  year={2021},
  publisher={IEEE}
}

@inproceedings{zhang2023mitigating,
  title={Mitigating persistence of open-source vulnerabilities in maven ecosystem},
  author={Zhang, Lyuye and Liu, Chengwei and Chen, Sen and Xu, Zhengzi and Fan, Lingling and Zhao, Lida and Zhang, Yiran and Liu, Yang},
  booktitle={2023 38th IEEE/ACM International Conference on Automated Software Engineering (ASE)},
  pages={191--203},
  year={2023},
  organization={IEEE}
}

@inproceedings{shi2022backporting,
  title={Backporting security patches of web applications: A prototype design and implementation on injection vulnerability patches},
  author={Shi, Youkun and Zhang, Yuan and Luo, Tianhan and Mao, Xiangyu and Cao, Yinzhi and Wang, Ziwen and Zhao, Yudi and Huang, Zongan and Yang, Min},
  booktitle={31st USENIX Security Symposium (USENIX Security 22)},
  pages={1993--2010},
  year={2022}
}

@article{shariffdeen2020automated,
  title={Automated patch transplantation},
  author={Shariffdeen, Ridwan Salihin and Tan, Shin Hwei and Gao, Mingyuan and Roychoudhury, Abhik},
  journal={ACM Transactions on Software Engineering and Methodology (TOSEM)},
  volume={30},
  number={1},
  pages={1--36},
  year={2020},
  publisher={ACM New York, NY, USA}
}

@article{ren2020codebleu,
  title={Codebleu: a method for automatic evaluation of code synthesis},
  author={Ren, Shuo and Guo, Daya and Lu, Shuai and Zhou, Long and Liu, Shujie and Tang, Duyu and Sundaresan, Neel and Zhou, Ming and Blanco, Ambrosio and Ma, Shuai},
  journal={arXiv preprint arXiv:2009.10297},
  year={2020}
}

@inproceedings{ray2012case,
  title={A case study of cross-system porting in forked projects},
  author={Ray, Baishakhi and Kim, Miryung},
  booktitle={Proceedings of the ACM SIGSOFT 20th International Symposium on the Foundations of Software Engineering},
  pages={1--11},
  year={2012}
}

@inproceedings{chakroborti2024study,
  title={A Study of Backporting Code in Open-Source Software for Characterizing Changesets},
  author={Chakroborti, Debasish and Roy, Chanchal and Schneider, Kevin},
  booktitle={Proceedings of the 2024 IEEE/ACM 46th International Conference on Software Engineering: Companion Proceedings},
  pages={296--297},
  year={2024}
}

@article{li2025cocoevo,
  title={CoCoEvo: Co-Evolution of Programs and Test Cases to Enhance Code Generation},
  author={Li, Kefan and Yuan, Yuan and Yu, Hongyue and Guo, Tingyu and Cao, Shijie},
  journal={IEEE Transactions on Evolutionary Computation},
  year={2025},
  publisher={IEEE}
}

@article{he2025llm,
  title={LLM-Based Multi-Agent Systems for Software Engineering: Literature Review, Vision, and the Road Ahead},
  author={He, Junda and Treude, Christoph and Lo, David},
  journal={ACM Transactions on Software Engineering and Methodology},
  volume={34},
  number={5},
  pages={1--30},
  year={2025},
  publisher={ACM New York, NY}
}

@misc{openai2024swe,
  author = {OpenAI},
  title = {Introducing SWE-bench Verified
},
  year = 2024,
  url = {https://openai.com/index/introducing-swe-bench-verified/},
  lastaccessed = {August 21, 2025}
}

@article{zan2024swe,
  title={Swe-bench-java: A github issue resolving benchmark for java},
  author={Zan, Daoguang and Huang, Zhirong and Yu, Ailun and Lin, Shaoxin and Shi, Yifan and Liu, Wei and Chen, Dong and Qi, Zongshuai and Yu, Hao and Yu, Lei and others},
  journal={arXiv preprint arXiv:2408.14354},
  year={2024}
}

@article{monperrus2018automatic,
  title={Automatic software repair: A bibliography},
  author={Monperrus, Martin},
  journal={ACM Computing Surveys (CSUR)},
  volume={51},
  number={1},
  pages={1--24},
  year={2018},
  publisher={ACM New York, NY, USA}
}

@inproceedings{jiang2018shaping,
  title={Shaping program repair space with existing patches and similar code},
  author={Jiang, Jiajun and Xiong, Yingfei and Zhang, Hongyu and Gao, Qing and Chen, Xiangqun},
  booktitle={Proceedings of the 27th ACM SIGSOFT international symposium on software testing and analysis},
  pages={298--309},
  year={2018}
}

@article{xu2024aligning,
  title={Aligning llms for fl-free program repair},
  author={Xu, Junjielong and Fu, Ying and Tan, Shin Hwei and He, Pinjia},
  journal={CoRR},
  year={2024}
}

@article{chen2021evaluating,
  title={Evaluating large language models trained on code},
  author={Chen, Mark and Tworek, Jerry and Jun, Heewoo and Yuan, Qiming and Pinto, Henrique Ponde De Oliveira and Kaplan, Jared and Edwards, Harri and Burda, Yuri and Joseph, Nicholas and Brockman, Greg and others},
  journal={arXiv preprint arXiv:2107.03374},
  year={2021}
}

@article{zhuo2024bigcodebench,
  title={Bigcodebench: Benchmarking code generation with diverse function calls and complex instructions},
  author={Zhuo, Terry Yue and Vu, Minh Chien and Chim, Jenny and Hu, Han and Yu, Wenhao and Widyasari, Ratnadira and Yusuf, Imam Nur Bani and Zhan, Haolan and He, Junda and Paul, Indraneil and others},
  journal={arXiv preprint arXiv:2406.15877},
  year={2024}
}

@inproceedings{galenson2014codehint,
  title={Codehint: Dynamic and interactive synthesis of code snippets},
  author={Galenson, Joel and Reames, Philip and Bodik, Rastislav and Hartmann, Bj{\"o}rn and Sen, Koushik},
  booktitle={Proceedings of the 36th International Conference on Software Engineering},
  pages={653--663},
  year={2014}
}

@article{robertson2009probabilistic,
  title={The probabilistic relevance framework: BM25 and beyond},
  author={Robertson, Stephen and Zaragoza, Hugo and others},
  journal={Foundations and Trends{\textregistered} in Information Retrieval},
  volume={3},
  number={4},
  pages={333--389},
  year={2009},
  publisher={Now Publishers, Inc.}
}

@inproceedings{feldt2023towards,
  title={Towards autonomous testing agents via conversational large language models},
  author={Feldt, Robert and Kang, Sungmin and Yoon, Juyeon and Yoo, Shin},
  booktitle={2023 38th IEEE/ACM International Conference on Automated Software Engineering (ASE)},
  pages={1688--1693},
  year={2023},
  organization={IEEE}
}

@article{liu2024large,
  title={Large language model-based agents for software engineering: A survey},
  author={Liu, Junwei and Wang, Kaixin and Chen, Yixuan and Peng, Xin and Chen, Zhenpeng and Zhang, Lingming and Lou, Yiling},
  journal={arXiv preprint arXiv:2409.02977},
  year={2024}
}

@article{wang2025agents,
  title={Agents in software engineering: Survey, landscape, and vision},
  author={Wang, Yanlin and Zhong, Wanjun and Huang, Yanxian and Shi, Ensheng and Yang, Min and Chen, Jiachi and Li, Hui and Ma, Yuchi and Wang, Qianxiang and Zheng, Zibin},
  journal={Automated Software Engineering},
  volume={32},
  number={2},
  pages={1--36},
  year={2025},
  publisher={Springer}
}

@article{zhang2024large,
  title={Large language model-brained gui agents: A survey},
  author={Zhang, Chaoyun and He, Shilin and Qian, Jiaxu and Li, Bowen and Li, Liqun and Qin, Si and Kang, Yu and Ma, Minghua and Liu, Guyue and Lin, Qingwei and others},
  journal={arXiv preprint arXiv:2411.18279},
  year={2024}
}

@article{dong2025survey,
  title={A Survey on Code Generation with LLM-based Agents},
  author={Dong, Yihong and Jiang, Xue and Qian, Jiaru and Wang, Tian and Zhang, Kechi and Jin, Zhi and Li, Ge},
  journal={arXiv preprint arXiv:2508.00083},
  year={2025}
}

@article{yang2025survey,
  title={A Survey of LLM-based Automated Program Repair: Taxonomies, Design Paradigms, and Applications},
  author={Yang, Boyang and Cai, Zijian and Liu, Fengling and Le, Bach and Zhang, Lingming and Bissyand{\'e}, Tegawend{\'e} F and Liu, Yang and Tian, Haoye},
  journal={arXiv preprint arXiv:2506.23749},
  year={2025}
}

@inproceedings{zhang2024autocoderover,
  title={Autocoderover: Autonomous program improvement},
  author={Zhang, Yuntong and Ruan, Haifeng and Fan, Zhiyu and Roychoudhury, Abhik},
  booktitle={Proceedings of the 33rd ACM SIGSOFT International Symposium on Software Testing and Analysis},
  pages={1592--1604},
  year={2024}
}

@article{antoniades2024swe,
  title={Swe-search: Enhancing software agents with monte carlo tree search and iterative refinement},
  author={Antoniades, Antonis and {\"O}rwall, Albert and Zhang, Kexun and Xie, Yuxi and Goyal, Anirudh and Wang, William},
  journal={arXiv preprint arXiv:2410.20285},
  year={2024}
}

@article{yang2025enhancing,
  title={Enhancing Repository-Level Software Repair via Repository-Aware Knowledge Graphs},
  author={Yang, Boyang and Tian, Haoye and Ren, Jiadong and Jin, Shunfu and Liu, Yang and Liu, Feng and Le, Bach},
  journal={arXiv preprint arXiv:2503.21710},
  year={2025}
}

@article{wang2024openhands,
  title={Openhands: An open platform for ai software developers as generalist agents},
  author={Wang, Xingyao and Li, Boxuan and Song, Yufan and Xu, Frank F and Tang, Xiangru and Zhuge, Mingchen and Pan, Jiayi and Song, Yueqi and Li, Bowen and Singh, Jaskirat and others},
  journal={arXiv preprint arXiv:2407.16741},
  year={2024}
}

@article{su2025learn,
  title={Learn-by-interact: A data-centric framework for self-adaptive agents in realistic environments},
  author={Su, Hongjin and Sun, Ruoxi and Yoon, Jinsung and Yin, Pengcheng and Yu, Tao and Ar{\i}k, Sercan {\"O}},
  journal={arXiv preprint arXiv:2501.10893},
  year={2025}
}

@article{gao2025trae,
  title={Trae Agent: An LLM-based Agent for Software Engineering with Test-time Scaling},
  author={Gao, Pengfei and Tian, Zhao and Meng, Xiangxin and Wang, Xinchen and Hu, Ruida and Xiao, Yuanan and Liu, Yizhou and Zhang, Zhao and Chen, Junjie and Gao, Cuiyun and others},
  journal={arXiv preprint arXiv:2507.23370},
  year={2025}
}

@article{tao2024magis,
  title={Magis: Llm-based multi-agent framework for github issue resolution},
  author={Tao, Wei and Zhou, Yucheng and Wang, Yanlin and Zhang, Wenqiang and Zhang, Hongyu and Cheng, Yu},
  journal={Advances in Neural Information Processing Systems},
  volume={37},
  pages={51963--51993},
  year={2024}
}

@misc{OSV,
  author = {OSV Team},
  title = {A distributed vulnerability database for Open Source},
  year = 2025,
  url = {https://osv.dev/},
  lastaccessed = {August 21, 2025}
}

@misc{Chang2025aliases,
  author = {Chang, Oliver and Cox, Russ},
  title = {Open Source Vulnerability format},
  year = 2025,
  url = {https://ossf.github.io/osv-schema/#aliases-field},
  lastaccessed = {August 21, 2025}
}

@misc{NVD,
  title = {National Vulnerability Database},
  year = 2025,
  url = {https://nvd.nist.gov/},
  lastaccessed = {August 21, 2025}
}

@misc{GAD,
  title = {GitHub Advisory Database: Security vulnerability database inclusive of CVEs and GitHub originated security advisories from the world of open source software.},
  year = 2025,
  url = {https://github.com/github/advisory-database/},
  lastaccessed = {August 21, 2025}
}

@inproceedings{zahan2023software,
  title={Do software security practices yield fewer vulnerabilities?},
  author={Zahan, Nusrat and Shohan, Shohanuzzaman and Harris, Dan and Williams, Laurie},
  booktitle={2023 IEEE/ACM 45th International Conference on Software Engineering: Software Engineering in Practice (ICSE-SEIP)},
  pages={292--303},
  year={2023},
  organization={IEEE}
}

@inproceedings{yang2025tracing,
  title={Tracing vulnerabilities in maven: a study of cve lifecycles and dependency networks},
  author={Yang-Smith, Corey and Abdellatif, Ahmad},
  booktitle={2025 IEEE/ACM 22nd International Conference on Mining Software Repositories (MSR)},
  pages={349--353},
  year={2025},
  organization={IEEE}
}

@inproceedings{pan2024towards,
  title={Towards more practical automation of vulnerability assessment},
  author={Pan, Shengyi and Bao, Lingfeng and Zhou, Jiayuan and Hu, Xing and Xia, Xin and Li, Shanping},
  booktitle={Proceedings of the IEEE/ACM 46th International Conference on Software Engineering},
  pages={1--13},
  year={2024}
}

@article{nashid2025issue2test,
  title={Issue2Test: Generating Reproducing Test Cases from Issue Reports},
  author={Nashid, Noor and Bouzenia, Islem and Pradel, Michael and Mesbah, Ali},
  journal={arXiv preprint arXiv:2503.16320},
  year={2025}
}

@article{hu2024real,
  title={A Real-World Benchmark for Evaluating Fine-Grained Issue Solving Capabilities of Large Language Models},
  author={Hu, Ruida and Peng, Chao and Ren, Jingyi and Jiang, Bo and Meng, Xiangxin and Wu, Qinyun and Gao, Pengfei and Wang, Xinchen and Gao, Cuiyun},
  journal={arXiv preprint arXiv:2411.18019},
  year={2024}
}

@article{mundler2024swt,
  title={SWT-bench: Testing and validating real-world bug-fixes with code agents},
  author={M{\"u}ndler, Niels and M{\"u}ller, Mark and He, Jingxuan and Vechev, Martin},
  journal={Advances in Neural Information Processing Systems},
  volume={37},
  pages={81857--81887},
  year={2024}
}

@article{venturini2023depended,
  title={I depended on you and you broke me: An empirical study of manifesting breaking changes in client packages},
  author={Venturini, Daniel and Cogo, Filipe Roseiro and Polato, Ivanilton and Gerosa, Marco A and Wiese, Igor Scaliante},
  journal={ACM Transactions on Software Engineering and Methodology},
  volume={32},
  number={4},
  pages={1--26},
  year={2023},
  publisher={ACM New York, NY, USA}
}

@inproceedings{jayasuriya2023understanding,
  title={Understanding breaking changes in the wild},
  author={Jayasuriya, Dhanushka and Terragni, Valerio and Dietrich, Jens and Ou, Samuel and Blincoe, Kelly},
  booktitle={Proceedings of the 32nd ACM SIGSOFT International Symposium on Software Testing and Analysis},
  pages={1433--1444},
  year={2023}
}

@inproceedings{pashchenko2020qualitative,
  title={A qualitative study of dependency management and its security implications},
  author={Pashchenko, Ivan and Vu, Duc-Ly and Massacci, Fabio},
  booktitle={Proceedings of the 2020 ACM SIGSAC conference on computer and communications security},
  pages={1513--1531},
  year={2020}
}

@inproceedings{wingerd1998high,
  title={High-level best practices in software configuration management},
  author={Wingerd, Laura and Seiwald, Christopher},
  booktitle={International Workshop on Software Configuration Management},
  pages={57--66},
  year={1998},
  organization={Springer}
}

@inproceedings{shihab2012effect,
  title={The effect of branching strategies on software quality},
  author={Shihab, Emad and Bird, Christian and Zimmermann, Thomas},
  booktitle={Proceedings of the ACM-IEEE international symposium on Empirical software engineering and measurement},
  pages={301--310},
  year={2012}
}

@inproceedings{tan2022understanding,
  title={Understanding the practice of security patch management across multiple branches in oss projects},
  author={Tan, Xin and Zhang, Yuan and Cao, Jiajun and Sun, Kun and Zhang, Mi and Yang, Min},
  booktitle={Proceedings of the ACM Web Conference 2022},
  pages={767--777},
  year={2022}
}

@inproceedings{rodriguez2015increasing,
  title={Increasing automation in the backporting of Linux drivers using Coccinelle},
  author={Rodriguez, Luis R and Lawall, Julia},
  booktitle={2015 11th European Dependable Computing Conference (EDCC)},
  pages={132--143},
  year={2015},
  organization={IEEE}
}

@inproceedings{chakroborti2022backports,
  title={Backports: Change types, challenges and strategies},
  author={Chakroborti, Debasish and Schneider, Kevin A and Roy, Chanchal K},
  booktitle={Proceedings of the 30th IEEE/ACM International Conference on Program Comprehension},
  pages={636--647},
  year={2022}
}

@misc{osv2025openssf,
  author = {OSV Team},
  title = {OpenSSF Malicious Packages
},
  year = 2025,
  url = {https://github.com/ossf/malicious-packages},
  lastaccessed = {September 5, 2025}
}

@misc{nvd2025vulnerability,
  author = {NVD},
  title = {NVD - Vulnerability Metrics
},
  year = 2025,
  url = {https://nvd.nist.gov/vuln-metrics/cvss},
  lastaccessed = {September 5, 2025}
}

@misc{SemVer,
  author = {Preston-Werner, Tom},
  title = {Semantic Versioning 2.0.0
},
  year = 2025,
  url = {https://semver.org/},
  lastaccessed = {September 5, 2025}
}

@inproceedings{zhong2025empirical,
  title={An Empirical Study on Package-Level Deprecation in Python Ecosystem},
  author={Zhong, Zhiqing and He, Shilin and Wang, Haoxuan and Yu, Boxi and Yang, Haowen and He, Pinjia},
  booktitle={2025 IEEE/ACM 47th International Conference on Software Engineering (ICSE)},
  pages={66--77},
  year={2025},
  organization={IEEE}
}

@misc{openai2025gp5,
  author = {OpenAI},
  title = {Introducing GPT-5
},
  year = 2025,
  url = {https://openai.com/index/introducing-gpt-5/},
  lastaccessed = {September 10, 2025}
}

@misc{anthropic2025sonnet4,
  author = {Anthropic},
  title = {Claude Sonnet 4
},
  year = 2025,
  url = {https://www.anthropic.com/claude/sonnet},
  lastaccessed = {September 10, 2025}
}

@misc{qwen3technicalreport,
      title={Qwen3 Technical Report}, 
      author={Qwen Team},
      year={2025},
      eprint={2505.09388},
      archivePrefix={arXiv},
      primaryClass={cs.CL},
      url={https://arxiv.org/abs/2505.09388}, 
}

@article{guo2025swe,
  title={SWE-Factory: Your Automated Factory for Issue Resolution Training Data and Evaluation Benchmarks},
  author={Guo, Lianghong and Wang, Yanlin and Li, Caihua and Yang, Pengyu and Chen, Jiachi and Tao, Wei and Zou, Yingtian and Tang, Duyu and Zheng, Zibin},
  journal={arXiv preprint arXiv:2506.10954},
  year={2025}
}

@article{zhang2025swe,
  title={SWE-bench Goes Live!},
  author={Zhang, Linghao and He, Shilin and Zhang, Chaoyun and Kang, Yu and Li, Bowen and Xie, Chengxing and Wang, Junhao and Wang, Maoquan and Huang, Yufan and Fu, Shengyu and others},
  journal={arXiv preprint arXiv:2505.23419},
  year={2025}
}

@misc{cass2025top,
  author = {Cass, Stephen},
  title = {The Top Programming Language 2024
},
  year = 2024,
  url = {https://spectrum.ieee.org/top-programming-languages-2024},
  lastaccessed = {September 11, 2025}
}

@article{wang2023plumber,
  title={Plumber: Boosting the propagation of vulnerability fixes in the npm ecosystem},
  author={Wang, Ying and Sun, Peng and Pei, Lin and Yu, Yue and Xu, Chang and Cheung, Shing-Chi and Yu, Hai and Zhu, Zhiliang},
  journal={IEEE Transactions on Software Engineering},
  volume={49},
  number={5},
  pages={3155--3181},
  year={2023},
  publisher={IEEE}
}

@article{cheng2025codemenv,
  title={Codemenv: Benchmarking large language models on code migration},
  author={Cheng, Keyuan and Shen, Xudong and Yang, Yihao and Wang, Tengyue and Cao, Yang and Ali, Muhammad Asif and Wang, Hanbin and Hu, Lijie and Wang, Di},
  journal={arXiv preprint arXiv:2506.00894},
  year={2025}
}

@article{liu2025migrationbench,
  title={MigrationBench: Repository-Level Code Migration Benchmark from Java 8},
  author={Liu, Linbo and Liu, Xinle and Zhou, Qiang and Chen, Lin and Liu, Yihan and Nguyen, Hoan and Omidvar-Tehrani, Behrooz and Shen, Xi and Huan, Jun and Tripp, Omer and others},
  journal={arXiv preprint arXiv:2505.09569},
  year={2025}
}

@inproceedings{xie2024unveiling,
  title={Unveiling the Characteristics and Impact of Security Patch Evolution},
  author={Xie, Zifan and Wen, Ming and Wei, Zichao and Jin, Hai},
  booktitle={Proceedings of the 39th IEEE/ACM International Conference on Automated Software Engineering},
  pages={1094--1106},
  year={2024}
}

\end{document}